\documentclass[12pt, a4paper]{article}
\usepackage{epsf}
\usepackage{cite}
\usepackage{amsmath,amssymb}
\input{colordvi.tex}
\usepackage[usenames,dvipsnames]{color}
\usepackage{graphicx}
\usepackage{ascmac}
\usepackage{hyperref}
\usepackage{slashed}
\usepackage{tikz-feynman}
\tikzfeynmanset{compat=1.0.0}

\begin{document}

\begin{titlepage}

\begin{center}
\hfill TU-1291\\
\hfill KEK-QUP-2025-0028
\vskip 1.in

\renewcommand{\thefootnote}{\fnsymbol{footnote}}

{\Large \bf
High Frequency Spectrum of Primordial \\ \vspace{2.5mm}
Gravitational Waves 
}

\vskip .5in

{\large
Kamil Mudrunka$^{(a)}$
and
Kazunori Nakayama$^{(a,b)}$
}

\vskip 0.5in

$^{(a)}${\em 
Department of Physics, Tohoku University, Sendai 980-8578, Japan
}

\vskip 0.2in

$^{(b)}${\em 
International Center for Quantum-field Measurement Systems for Studies of the Universe and Particles (QUP), KEK, 1-1 Oho, Tsukuba, Ibaraki 305-0801, Japan
}

\end{center}
\vskip .5in

\begin{abstract}

During inflation long wavelength gravitational waves are produced, which form stochastic background in the present universe with  very wide range of frequencies.
Higher frequency gravitational waves never experience super-Hubble-radius regime, but they are also amplified after inflation due to the inflaton oscillation.
Taking account of the inflaton dynamics after inflation, the spectrum of such quantum-mechanically produced gravitational waves may extend to much higher frequencies than previously thought.
In this paper we calculate the spectrum of high frequency gravitational waves produced during and after inflation in detail, in particular focusing on the connection between the low and high frequency regime, and show that the detailed spectrum can distinguish inflation models.

\end{abstract}

\end{titlepage}

\tableofcontents

\renewcommand{\thefootnote}{\arabic{footnote}}
\setcounter{footnote}{0}

\section{Introduction}

During inflation~\cite{Starobinsky:1980te,Guth:1980zm,Sato:1981qmu,Kazanas:1980tx,Linde:1981mu,Albrecht:1982wi}, scalar and tensor perturbations are stretched over the Hubble radius.
Long wavelength scalar perturbations generated during inflation are considered to be the origin of the structure of the universe.
On the other hand, long wavelength tensor perturbations form the so-called primordial gravitational waves (GWs), which result in the stochastic GW background in the present universe with various frequencies, ranging from cosmological scales to terrestrial scales~\cite{Starobinsky:1979ty,Allen:1987bk,Turner:1990rc,Turner:1993vb,Turner:1996ck,Smith:2005mm,Boyle:2005se}.
If primordial GWs will be detected in future, it provides us with the information about the inflation energy scale.

\begin{figure}
\centering
\includegraphics[width=0.8\textwidth]{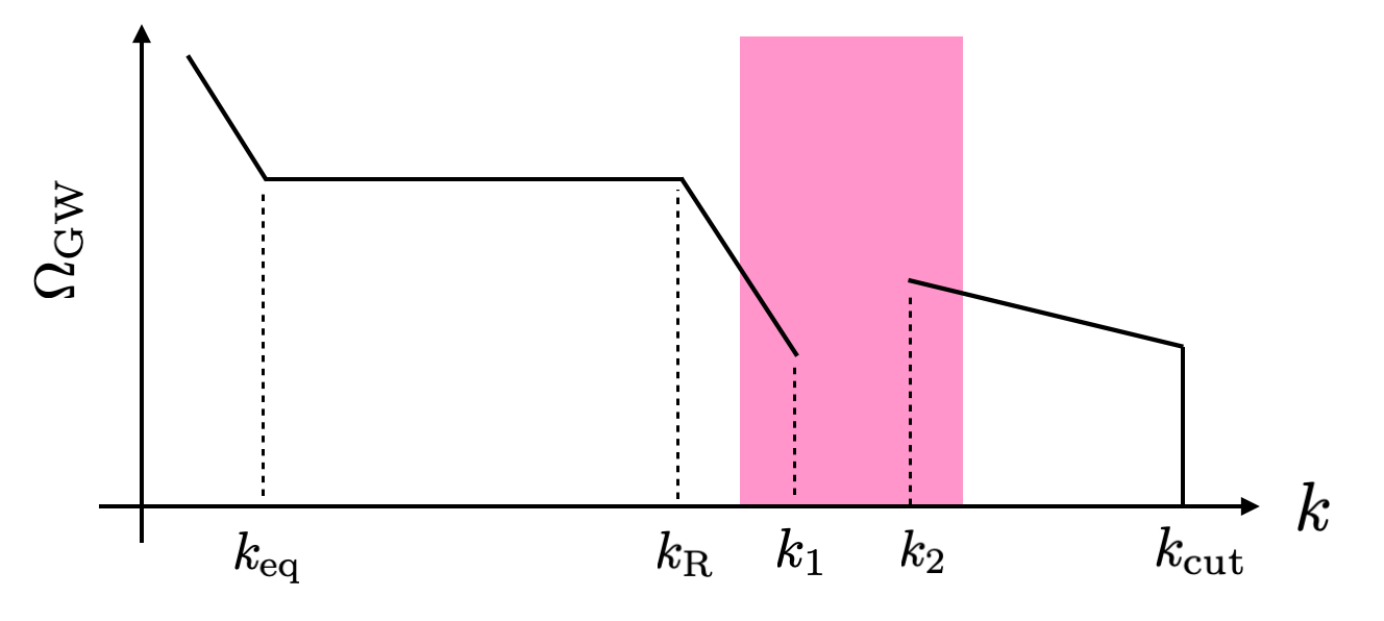}
\caption{
Schematic picture of the primordial GW spectrum for the case of $w=0$. 
Several characteristic comoving wavenumber are plotted: $k_{\rm eq}$ and $k_{\rm R}$ are the comoving Hubble scale at the matter-radiation equality and the end of reheating, respectively, and $k_{\rm cut}$ is the comoving inflaton mass scale at the end of reheating. For $k_1$ and $k_2$, see text. Our main focus is the behavior around the red shaded region.
}
\label{fig:gwbg}
\end{figure}

Not only the inflation energy scale, but also the information about thermal history (or equation of state of the universe, more precisely) of the universe after inflation is contained in the primordial GW spectrum~\cite{Seto:2003kc,Tashiro:2003qp,Nakayama:2008ip,Nakayama:2008wy,Kuroyanagi:2008ye,Mukohyama:2009zs,Nakayama:2009ce,Durrer:2011bi,Jinno:2011sw,Kuroyanagi:2011fy,Jinno:2012xb,Jinno:2013xqa,Jinno:2014qka,Kuroyanagi:2014qza,Minami:2025waa}.
If the equation of state parameter of the universe is $w$ when a super-Hubble-radius GW mode with comoving wavenumber $k$ re-enters the horizon, the resulting GW spectrum scales as $\Omega_{\rm GW}\propto k^{2(3w-1)/(1+3w)}$~\cite{Tashiro:2003qp,Nakayama:2008wy,Mukohyama:2009zs} for the scale-invariant long-wavelength tensor power spectrum (see Eq.~(\ref{OGW_ana_low})).
Therefore, the spectral shape of primordial GWs directly tells us equation of state of the universe.
Note that these arguments apply to the GW modes that once experienced super-Hubble-radius regime during inflation, i.e., $k \lesssim a_{\rm e} H_{\rm e} \equiv k_1$ where $a_{\rm e}$ and $H_{\rm e}$ are the scale factor and the Hubble scale at the end of inflation, respectively.
A schematic picture of the GW spectrum for $w=0$ is shown in Fig.~\ref{fig:gwbg}: the $k<k_1$ part corresponds to the modes that experienced super-Hubble-radius regime during inflation.
The flat part in the region $k_{\rm eq}<k<k_{\rm R}$, where $k_{\rm eq}$ and $k_{\rm R}$ are the comoving Hubble scale at the matter-radiation equality and the end of reheating, corresponds to the mode that re-enter the horizon during radiation-dominated era.

Recently it has been pointed out in Refs.~\cite{Ema:2015dka,Ema:2016hlw,Ema:2020ggo} that even higher frequency modes $k\gg k_1$ may be excited due to the post-inflationary inflaton dynamics.\footnote{
    It has been long known that the preheating dynamics can also produces high frequency GWs~\cite{Khlebnikov:1997di,Easther:2006vd,Easther:2006gt,Garcia-Bellido:2007nns,Garcia-Bellido:2007fiu,Dufaux:2007pt,Dufaux:2008dn}. Such GWs are produced by a source term in the GW equation of motion, indicating that the production is essentially \textit{classical}. On the other hand, what we are considering now is purely \textit{quantum} production, in a sense that GWs are produced without any source term.}
This is because the inflaton exhibits coherent oscillation and there appears an inflaton mass scale $m_\phi$, which is typically (much) larger than the Hubble scale, after inflation ends.\footnote{
    Here $m_\phi$ corresponds to the coefficient of the quadratic term when expanding the inflaton potential as $V(\phi)\simeq \frac{1}{2}m_\phi^2\varphi^2$ for $\phi=v+\varphi$ with $v$ being the field value at the potential minimum. If there does not exist such a quadratic term, we should instead define the field-dependent effective mass.}
This is nothing but particle production due to the rapidly oscillating background~\cite{Dolgov:1989us,Traschen:1990sw,Shtanov:1994ce,Kofman:1994rk,Kofman:1997yn} through the gravitational interaction, and can intuitively be regarded as inflaton annihilation to the graviton pair~\cite{Ema:2015dka,Ema:2016hlw,Ema:2020ggo,Choi:2024ilx,Xu:2024fjl,Bernal:2025lxp,Xu:2025wjq}.
This may result in a high frequency tail in the GW spectrum that extends to $k\gg k_1$.\footnote{
    In this paper we also call such high frequency GWs ``primordial GWs''.
    It is partly because the production of both low and high frequency modes are governed by exactly the same GW equation of motion under the homogeneous inflaton dynamics.
    Thus they are indistinguishable in rigorous sense, and actually the transition from the low to high frequency regime is smooth. 
}
If the inflaton oscillation proceeds through quadratic inflaton potential, i.e. $w=0$, such high frequency tail scales as $k^{-1/2}$ for $k\gg k_2 \equiv a_{\rm e} m_\phi$~\cite{Ema:2015dka,Ema:2020ggo}.
This contribution is schematically shown in Fig.~\ref{fig:gwbg}: the high frequency spectrum extends to $k>k_2$ and there is eventually a cutoff at $k=k_{\rm cut}$, the comoving inflaton mass scale at the end of reheating.\footnote{
    For $w>1/3$, the contribution from inflaton oscillation extends toward \textit{lower} frequency~\cite{Choi:2024ilx}, as we will explain in Sec.~\ref{sec:quantum}.
}

Therefore, the primordial GW spectrum for $k\lesssim k_1$ and $k\gtrsim k_2$ are relatively well understood.
However, depending on inflation models, there is large hierarchy between $k_1$ and $k_2$ and the spectrum at intermediate frequency range $k_1 \lesssim k \lesssim k_2$ has not been explored. 
The purpose of this paper is to precisely calculate the GW spectrum across the intermediate frequency range, as shown by the red shaded region in Fig.~\ref{fig:gwbg}.
Technically, the GW spectrum at low frequency limit is easily calculated in a classical way, while the Bogoliubov method is efficient for calculating the high frequency limit.
The intermediate frequency range requires careful treatment, and we will develop methods that can be applied to such a frequency range.

This paper is organized as follows.
In Sec.~\ref{sec:quantum} we give a formalism to calculate the primordial GW spectrum.
In Sec.~\ref{sec:GW} we calculate the GW spectrum for several concrete inflation models.
Sec.~\ref{sec:conc} is devoted to conclusions and discussion.

\section{Quantum production of gravitational waves}\label{sec:quantum}

\subsection{Formalism}

The GW or graviton can be identified as the tensor perturbation of the metric.
We define the tensor perturbation $h_{ij}$ around the Friedmann-Robertson-Walker metric as (see App.~\ref{sec:conv} for our convention)
\begin{align}
	ds^2=-dt^2 + a^2(t)(\delta_{ij} + h_{ij}(t,\vec x)) dx^idx^j,
\end{align}
where $a(t)$ is the scale factor. Let us Fourier expand $h_{ij}$ in terms of the creation-annihilation operators as
\begin{align}
	h_{ij}(t,\vec x) = \sum_{\lambda=+,\times} \int\frac{d^3k}{(2\pi)^3} \left[ a_{k,\lambda} h_{k,\lambda}(t) + a^\dagger_{-k,\lambda} h^*_{k,\lambda}(t)  \right]e^{i\vec k\cdot\vec x} e^\lambda_{ij},
	\label{hij_expand}
\end{align}
where the polarization tensor satisfies $e_{ij}^\lambda e_{ij}^{*\lambda'} =  \delta_{\lambda\lambda'}$.
The creation-annihilation operators satisfy the commutation relation
\begin{align}
	[ a_{k,\lambda}, a^\dagger_{k',\lambda'}  ] = (2\pi)^3 \delta(\vec k-\vec k') \delta_{\lambda\lambda'}.
\end{align}
We assume the standard Einstein-Hilbert action for a gravity sector.
Then the equation of motion of the graviton is
\begin{align}
	\ddot h_{k,\lambda} + 3 H \dot h_{k,\lambda} + \left(\frac{k}{a}\right)^2 h_{k,\lambda} =0,
	\label{h_eom}
\end{align}
where the dot denotes the derivative with respect to time $t$ and $H = \dot a/a$ denotes the Hubble parameter.
By defining $\widetilde h_{k,\lambda} \equiv a h_{k,\lambda} $, the equation of motion is also written as
\begin{align}
	\widetilde h_{k,\lambda}'' + \omega_k^2 \widetilde h_{k,\lambda} = 0,~~~~~~
	 \omega_k^2 = k^2 - \frac{a''}{a},
\end{align}
where the prime denotes the derivative with respect to the conformal time $\tau = \int \frac{dt}{a}$.
The initial condition is given in the high frequency limit $k \gg aH$ as
\begin{align}
	\widetilde h_{k,\lambda}(\tau) = \frac{1}{\sqrt{2k}} e^{-ik\tau}.
	\label{h_ini}
\end{align}
In the opposite limit $k \ll aH$ during inflation, the asymptotic solution is
\begin{align}
	\widetilde h_{k,\lambda}(\tau) = i\frac{a H_{\rm inf}}{\sqrt{2} k^{3/2}},
	\label{h_inf}
\end{align}
where $H_{\rm inf}$ denotes the Hubble scale during inflation.\footnote{
	The Hubble scale slowly changes during inflation. In this paper we are mostly interested in the Hubble parameter around the end of inflation, which we denote by $H_{\rm e}$ hereafter. 
}

\subsubsection*{Semi-classical method}

A straightforward method to evaluate the produced GW energy density is just to solve the equation of motion (\ref{h_eom}) with initial condition (\ref{h_ini}), under the background inflaton motion:
\begin{align}
	&\ddot \phi + 3H \dot \phi + \frac{\partial V}{\partial \phi} = 0,  \label{phi_eom} \\
	&3H^2 M_{\rm Pl}^2 = \frac{\dot \phi^2}{2} + V  \label{Fried},
\end{align}
where $\phi$ is the inflaton and $V$ is its scalar potential. Concrete inflation models will be listed in the following sections.
In this formalism, the Hubble parameter $H$ oscillates after inflation ends and it results in the graviton production~\cite{Ema:2015dka,Ema:2020ggo}.
The resulting GW energy density is given by
\begin{align}
	a^4(\tau)\rho_h(\tau) &= 2 \int \frac{d^3k}{(2\pi)^3}\frac{1}{2}\left[ \left| \widetilde h'_{k,\lambda}(\tau) \right|^2 + k^2 \left|\widetilde h_{k,\lambda}(\tau)\right|^2  -k\right] \\
	&\equiv\int d\ln k \, a^4(\tau) \rho_h(k, \tau) 
\end{align}
where
\begin{align}
	a^4(\tau) \rho_h(k, \tau) = \frac{k^3}{2\pi^2} \left[ \left| \widetilde h'_{k,\lambda}(\tau) \right|^2 + k^2 \left|\widetilde h_{k,\lambda}(\tau)\right|^2  -k\right].
	\label{rhoh_s}
\end{align}
Here the divergent zero-point energy is subtracted.
We call this the ``semi-classical method'', since we just need to solve the equation of motion (\ref{h_eom}) along with (\ref{phi_eom}) and (\ref{Fried}) to derive the resulting GW energy density, although the actual production process is purely quantum since we start from the zero-point fluctuation  (\ref{h_ini}) of the graviton field as the initial condition.

In principle this method works for arbitrary low and high graviton momenta, but for practical numerical calculation this method becomes more and more inefficient for high frequency limit $k \gg aH$ since we need extremely high precision to properly subtract the divergent part.
In the high frequency limit, the following Bogoliubov method is much more efficient.

\subsubsection*{Bogoliubov method}

In the Bogoliubov method, we further decompose $\widetilde h_{k,\lambda}(\tau)$ as
\begin{align}
	\widetilde h_{k,\lambda}(\tau) = \alpha_k(\tau) v_k(\tau) + \beta_k(\tau) v_k^*(\tau),~~~~~~v_k(\tau) \equiv  \frac{1}{\sqrt{2k}} e^{-i\Omega_k},
\end{align}
where $\alpha_k(\tau)$ and $\beta_k(\tau)$ are called Bogoliubov coefficient and $\Omega_k(\tau) = \int \omega_k(\tau) d\tau$.
They are chosen to satisfy the following relation:
\begin{align}
	\alpha'_k(\tau) = \frac{\omega_k'}{2\omega_k} \beta_k e^{2i\Omega_k},~~~~~~
	\beta'_k(\tau) = \frac{\omega_k'}{2\omega_k} \alpha_k e^{-2i\Omega_k}
	\label{eq_Bogo}
\end{align}
The initial condition (\ref{h_ini}) is written as $\alpha_k(\tau)=1$ and $\beta_k(\tau)=0$ in this language. One can easily prove that the relation $|\alpha_k(\tau)|^2-|\beta_k(\tau)|^2=1$ always holds.
This equation is also rewritten as
\begin{align}
	A_k ' = -i \omega_k A_k + \frac{\omega_k'}{2\omega_k} B_k,~~~~~~B_k ' = i \omega_k B_k + \frac{\omega_k'}{2\omega_k} A_k,
	\label{eq_Bogo'}
\end{align}
by defining $A_k \equiv \alpha_k e^{-i\Omega_k}$ and $B_k\equiv \beta_k e^{i\Omega_k}$.
For readers' convenience, we list some useful equations to numerically evaluate this equation:
\begin{align}
	&\frac{a''}{a} = a^2(2H^2 + \dot H), \\
	&\omega_k' = -\frac{a^3}{2\omega_k}\left(\ddot H + 6 H \dot H + 4 H^3 \right),\\
	&\dot H = -\frac{\dot\phi^2}{2M_{\rm Pl}^2},~~~~~~\ddot H = \frac{\dot\phi}{M_{\rm Pl}^2}\left( 3H \dot\phi + \frac{\partial V}{\partial \phi} \right).
\end{align}
One can integrate the equation (\ref{eq_Bogo}) or (\ref{eq_Bogo'}) along with the background evolution (\ref{phi_eom}) and (\ref{Fried}) to obtain the Bogoliubov coefficients at arbitrary later time.
One should note that $\omega_k$ becomes zero at the horizon crossing and the equation (\ref{eq_Bogo}) or (\ref{eq_Bogo'}) becomes singular.
Thus the Bogoliubov method is only useful for high frequency modes that never exit the horizon.

In terms of the Bogoliubov coefficients, the renormalized graviton energy density is given by~\cite{Birrell:1982ix,Ema:2018ucl}
\begin{align}
	a^4(\tau)\rho_h(\tau) &= 2 \int \frac{d^3k}{(2\pi)^3} \, k \, |\beta_k(\tau)|^2,
\end{align}
and hence
\begin{align}
	a^4(\tau) \rho_h(k, \tau) = \frac{k^4}{\pi^2} |\beta_k(\tau)|^2.
	\label{rhoh_B}
\end{align}
Thus $\beta_k$ counts only the physically produced particles. In this method we do not need to worry about the subtraction of the divergent part: it is already subtracted, since, by definition, the Bogoliubov method only counts the excitation around the vacuum (see e.g. Refs.~\cite{Shtanov:1994ce,Kofman:1997yn}).

\subsection{Gravitational wave spectrum}

Once we evaluate the GW energy density by Eqs.~(\ref{rhoh_s}) or (\ref{rhoh_B}) at some fixed cosmological time $t=t_{\rm c}$, it is converted to the present GW density parameter $\Omega_{\rm GW}(f)$. 
To do so, we need to assume background evolution of the universe.

Our basic assumption is that the inflaton oscillates around the potential minimum after inflation ends. 
Let us expand the inflaton field around its vacuum expectation value $v$ as $\phi = v + \varphi$ and suppose that the potential is approximated by $V \propto \varphi^n$ near the potential minimum.
The effective equation of state parameter $w$ during the inflaton oscillation regime is given by
\begin{align}
	w = \frac{n-2}{n+2},
\end{align}
and the inflaton energy density decreases as $\rho_\phi \propto a^{-3(1+w)} = a^{-\frac{6n}{n+2}}$.
Such an inflaton oscillation regime lasts for a while, and finally the inflaton is assumed to decay and reheat the universe.
After the decay, the universe enters the radiation-dominated regime.
The reheating temperature is represented by $T_{\rm R}$ hereafter.

With these assumptions, the present GW density parameter is given by
\begin{align}
	\Omega_{\rm GW}(f) &= \frac{\rho_h(k, t_0)}{\rho_{\rm crit}(t_0)} \\
	&= \Omega_r \times \frac{g_{*\rm R}}{g_{*0}} \left(\frac{g_{*s0}}{g_{*s\rm R}}\right)^{\frac{4}{3}}\times 
	\left(\frac{H_{\rm R}}{H_{\rm c}}\right)^{\frac{2(1-3w)}{3(1+w)}} \times \frac{\rho_h(k, t_{\rm c})}{3 H_{\rm c}^2 M_{\rm Pl}^2},
	\label{OGW_rhoh}
\end{align}
where $\Omega_r$ is the radiation density parameter and the subscript $0$, $\rm R$ and $\rm c$ represent the present time $t=t_0$, reheating time ($T=T_{\rm R}$) and $t=t_{\rm c}$, respectively.\footnote{
    Here $g_{*0}=3.36$, $g_{*s0}=3.909$ and $\Omega_r \simeq 8.5\times 10^{-5}$ are defined as if all neutrinos were massless.
    If one wants to avoid the confusion, one could use $g_{*}(T_{\rm eq}) =3.36$ and $g_{*s}(T_{\rm eq})=3.909$ at the matter-radiation equality instead of $g_{*0}$ and $g_{*s0}$, and the photon density parameter $\Omega_{\gamma}$ by replacing $\Omega_r = (g_{*0}/2)\Omega_{\gamma}$. 
}
The comoving wavenumber $k$ and the present GW frequency $f$ is related as
\begin{align}
	f = \frac{k}{2\pi a_0} &= \frac{k}{2\pi a_{\rm c}}\times \left(\frac{g_{*s0}}{g_{*s\rm R}}\right)^{\frac{1}{3}} \times \frac{T_0}{T_{\rm R}}\times \left(\frac{H_{\rm R}}{H_{\rm c}}\right)^{\frac{2}{3(1+w)}},\\
	&=  \frac{k}{2\pi a_{\rm R}}\times \left(\frac{g_{*s0}}{g_{*s\rm R}}\right)^{\frac{1}{3}} \times \frac{T_0}{T_{\rm R}},
\end{align}
where $T_0$ is the present cosmic temperature.
In the following sections, we assume $T_{\rm R}=10^{10}\,{\rm GeV}$ as a representative value and $g_{*\rm R}=g_{*s\rm R}=106.75$.

\subsection{Analytic estimation} \label{sec:rough}

\begin{figure}
\centering
\includegraphics[width=1.0\textwidth]{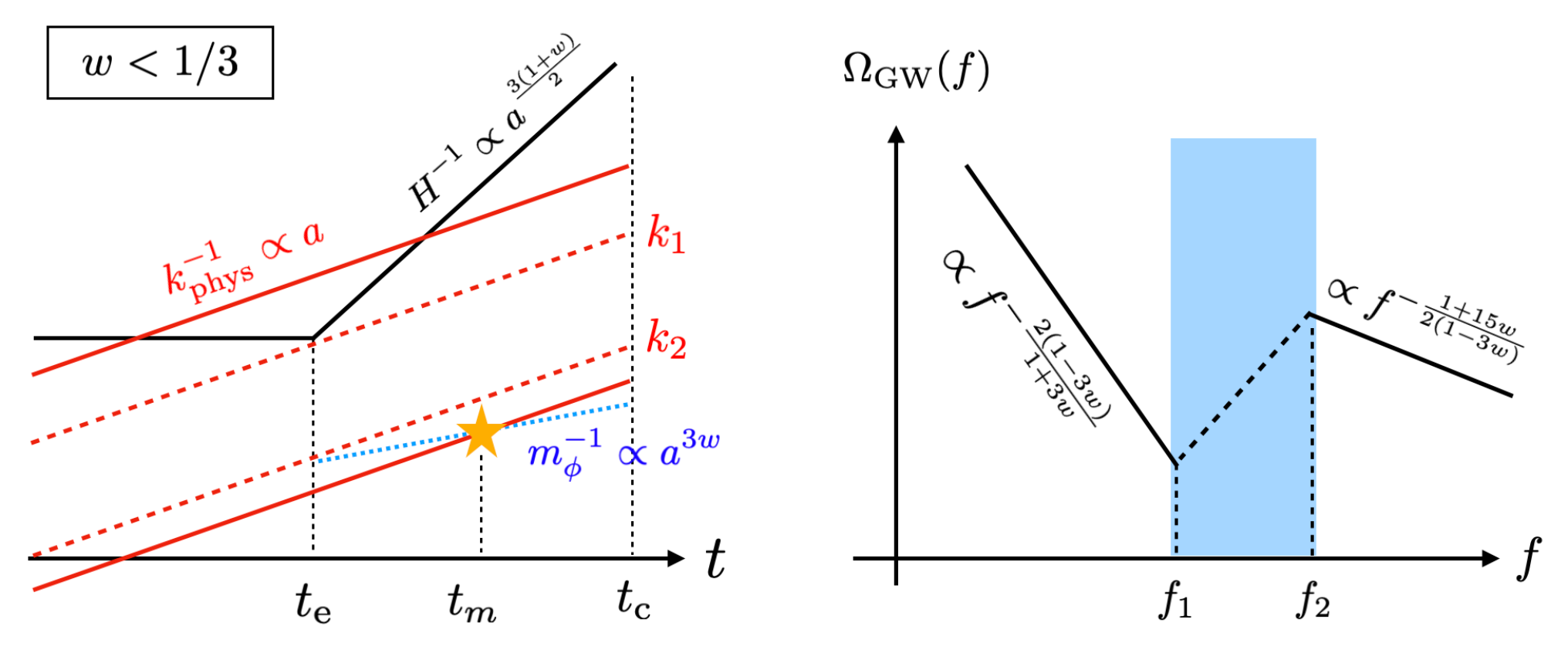}
\caption{(Left) Time evolution of several length scales: Hubble scale $H^{-1}$ (black line), physical GW wavelength $a/k$ (red lines), the effective inflaton mass scale $m_\phi^{-1}$ (blue dashed line). The time at the end of inflation is represented by $t_{\rm e}$ and the star represents when the condition  $k=am_\phi$ is satisfied for one choice of $k$ at $t=t_m$, at which particle production happens.
(Right) Expected GW spectrum as a function of (present) GW frequency $f$: $f_1$ and $f_2$ correspond to the comoving wavenumber $k_1$ and $k_2$. The blue shaded region is hard to predict analytically and we need numerical simulation to derive the behavior in this region.
We assumed $w<1/3$ in this figure, where $w$ is the equation of state parameter during the inflaton oscillation epoch.}
\label{fig:smallw}
\end{figure}
\begin{figure}
\centering
\includegraphics[width=1.0\textwidth]{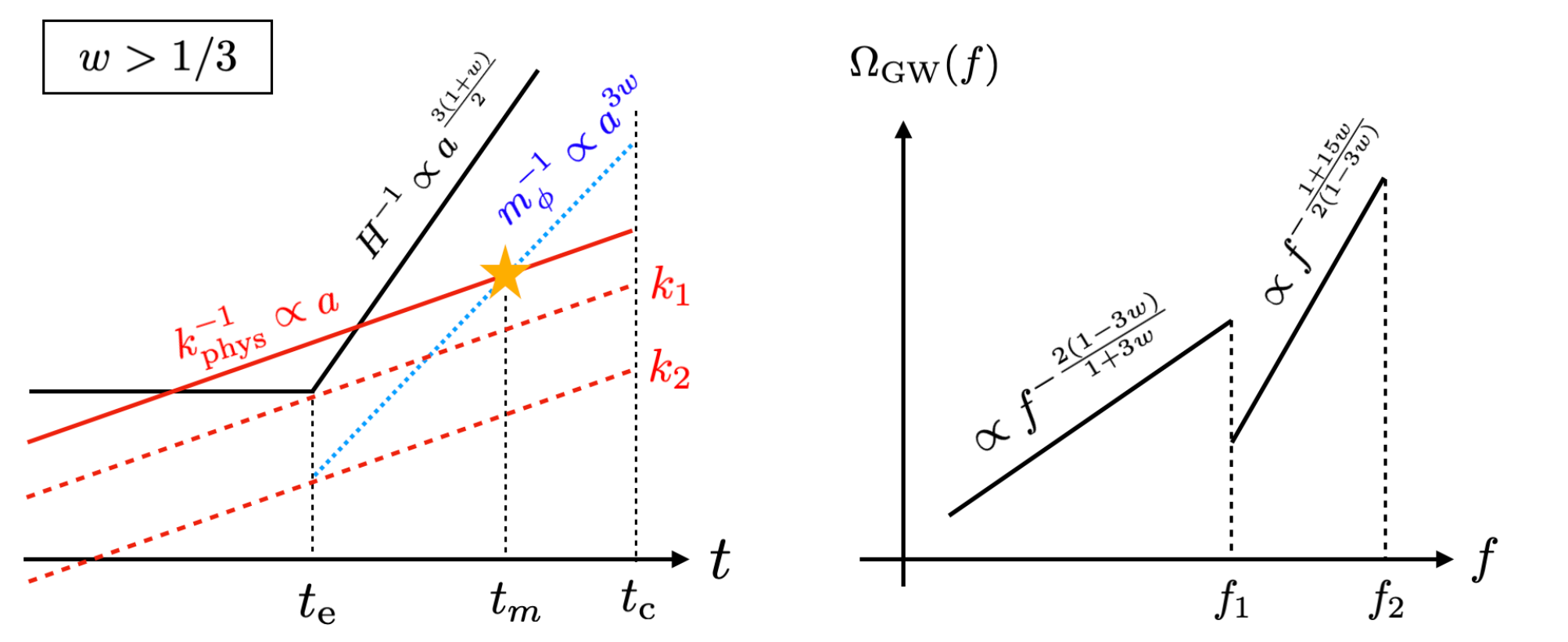}
\caption{The same as Fig.~\ref{fig:smallw}, but for $w>1/3$.}
\label{fig:largew}
\end{figure}

Here we give a rough estimation of the qualitative behavior of the GW spectrum.
Let us consider a typical situation shown in the left panel of Fig.~\ref{fig:smallw} for $w<1/3$ and Fig.~\ref{fig:largew} for $w>1/3$, in which the time evolution of the Hubble scale and several typical physical wavelength are shown. 
The time at which inflation ends is denoted by $t_{\rm e}$.

\subsubsection*{Long wavelength limit}

First we focus on the GW modes that once exit the Hubble radius during inflation, $k \ll a_{\rm e} H_{\rm e}$.
In this case, the GW amplitude is frozen at the value $\frac{k^3 h_k^2}{2\pi^2} \sim \left(\frac{H_{\rm e}}{2\pi}\right)^2\left(\frac{k}{k_*}\right)^{n_t}$, with $n_t$ being the tensor spectral index expressed in terms of the slow-roll parameter~\cite{Liddle:2000cg} and $k_*$ being some reference scale.
After it re-enters the horizon, the GW amplitude decreases as $h_k\propto a^{-1}$. 
Thus the GW energy density at $t=t_{\rm c}$ is given by
\begin{align}
	\rho_h(k, t_{\rm c}) \simeq \frac{k^3 h_k^2(t_{\rm c})}{2\pi^2} \left(\frac{k}{a_{\rm c}}\right)^2 \simeq 
	\left(\frac{H_{\rm e}}{2\pi}\frac{a_k}{a_c} \right)^2  \left(\frac{k}{a_{\rm c}}\right)^2
    \left(\frac{k}{k_*}\right)^{n_t},
\end{align}
where $a_k = a(t_k)$ with $t_k$ defined by the time at the horizon re-entry: $k = a(t_k) H(t_k)$. By using
\begin{align}
	\frac{a_k}{a_{\rm e}} = \left(\frac{a_{\rm e}H_{\rm e}}{k}\right)^{\frac{2}{1+3w}},
\end{align}
we obtain
\begin{align}
	\rho_h(k, t_{\rm c}) \simeq \frac{H_{\rm e}^4}{4\pi^2} \left(\frac{a_{\rm e}}{a_{\rm c}}\right)^4 \left(\frac{a_{\rm e}H_{\rm e}}{k}\right)^{\frac{2(1-3w)}{1+3w}}
    \left(\frac{k}{k_*}\right)^{n_t}.
	\label{rhoh_ana_low}
\end{align}
This reproduces the well known result that the GW spectrum is flat for the modes that enter the horizon at the radiation-dominated era $(w=1/3)$ and the spectrum scales as $k^{-2}$ for those which enter the horizon at the matter-dominated era $(w=0)$~\cite{Nakayama:2008ip,Nakayama:2008wy}.
In terms of $\Omega_{\rm GW}(f)$, with the use of formula (\ref{OGW_rhoh}), we obtain
\begin{align}
	\Omega_{\rm GW}(f) 
    &= \Omega_r \times \frac{g_{*\rm R}}{g_{*0}} \left(\frac{g_{*s0}}{g_{*s\rm R}}\right)^{\frac{4}{3}}\times 
	\frac{H_{\rm e}^2}{12\pi^2 M_{\rm Pl}^2}
    \left(\frac{H_{\rm R}}{H_{\rm e}}\right)^{\frac{2(1-3w)}{3(1+w)}}
	\left(\frac{a_{\rm e}H_{\rm e}}{k}\right)^{\frac{2(1-3w)}{1+3w}}
    \left(\frac{k}{k_*}\right)^{n_t}\\
    &= \Omega_r \times \frac{g_{*\rm R}}{g_{*0}} \left(\frac{g_{*s0}}{g_{*s\rm R}}\right)^{\frac{4}{3}}\times 
	\frac{H_{\rm e}^2}{12\pi^2 M_{\rm Pl}^2}
	\left(\frac{a_{\rm R}H_{\rm R}}{k}\right)^{\frac{2(1-3w)}{1+3w}}
    \left(\frac{k}{k_*}\right)^{n_t},
    \label{OGW_ana_low}
\end{align}
for $a_{\rm R}H_{\rm R} < k < a_{\rm e}H_{\rm e}$.
To obtain $\Omega_{\rm GW}$ for $k<a_{\rm R}H_{\rm R}$, one can just substitute $w=1/3$ in (\ref{OGW_ana_low}).
Note that the $t_{\rm c}$ dependence is gone as it should be.

\subsubsection*{Short wavelength limit}

Next let us consider GWs with short wavelength limit, which never exit the horizon.
As far as $k\gg a H_{\rm inf}$, the evolution of the modes are adiabatic during inflation and no enhancement happens.
However, after inflation, the inflaton coherent oscillation begins and then another mass scale, the inflaton mass $m_\phi$, appears.
It is sometimes much larger than the Hubble scale $H_{\rm inf}$.
Then the GW modes enhancement happens when the condition $k/a \simeq m_\phi$ is satisfied.
This is nothing but a particle production due to the oscillating field, and a simple (rough) interpretation is that the gravitons are produced by the inflaton annihilation process $\varphi\varphi \to hh$~\cite{Ema:2015dka,Ema:2020ggo}.
For the inflaton potential $V \sim \varphi^n$, the effective ``mass'' of the inflaton is given by $m_\phi \propto \overline\varphi^{(n-2)/2} \propto a^{-3w}$ (where $\overline\varphi$ is the amplitude of the oscillating field  $\varphi$) and there may also be multi inflaton annihilation processes like $\varphi \cdots \varphi \to hh$.

The effective production rate of the graviton from the inflaton is given by
\begin{align}
	\Gamma_{\varphi\varphi\to hh} \sim \mathcal C \frac{\overline\varphi^2 m_\phi^3 }{M_{\rm Pl}^4},
\end{align}
where $\mathcal C \sim 10^{-2}$ is a numerical factor~\cite{Ema:2015dka,Ema:2020ggo}.
Let us define the graviton production time $t_m$ so that it satisfies $k \simeq a(t_m) m_\phi(t_m)$.
The produced graviton energy density with the momentum $k$ is given by
\begin{align}
	\rho_h(k, t_{\rm c}) \sim \rho_\phi (t_m) \frac{\Gamma_{\varphi\varphi\to hh}}{H(t_m)} \left(\frac{a_m}{a_{\rm c}}\right)^4
	\sim \mathcal C\, m_\phi (t_m) H^3(t_m) \left(\frac{a_m}{a_{\rm c}}\right)^4.
\end{align}
By using
\begin{align}
	\frac{a_m}{a_{\rm e}} = \left(\frac{k}{a_{\rm e}m_\phi(t_{\rm e})}\right)^{\frac{1}{1-3w}},
\end{align}
we can rewrite the GW energy density as
\begin{align}
	\rho_h(k, t_{\rm c}) \sim \mathcal C\, m_\phi (t_{\rm e}) H_{\rm e}^3 \left(\frac{a_e}{a_{\rm c}}\right)^4 \left(\frac{a_{\rm e} m_\phi(t_{\rm e})}{k}\right)^{\frac{1+15w}{2(1-3w)}}.
	\label{rhoh_ana_high}
\end{align}
It reproduces the known result that the spectrum scales as $k^{-1/2}$ for constant inflaton mass $(w=0)$~\cite{Ema:2015dka,Ema:2020ggo}.
In terms of $\Omega_{\rm GW}(f)$, with the use of formula (\ref{OGW_rhoh}), we obtain
\begin{align}
	\Omega_{\rm GW}(f) = \Omega_r \times \frac{g_{*\rm R}}{g_{*0}} \left(\frac{g_{*s0}}{g_{*s\rm R}}\right)^{\frac{4}{3}}\times 
	\frac{\mathcal C m_\phi(t_{\rm e})H_{\rm e}}{3 M_{\rm Pl}^2}
	\left(\frac{H_{\rm R}}{H_{\rm e}}\right)^{\frac{2(1-3w)}{3(1+w)}} \left(\frac{a_{\rm e}m_\phi(t_{\rm e})}{k}\right)^{\frac{1+15w}{2(1-3w)}}.
    \label{OGW_ana_high}
\end{align}

Several remarks are in order. 
First, the expression (\ref{rhoh_ana_high}) is divergent for $w=1/3$ (or $n=4$). This represents the fact that $a(t) m_\phi(t) = {\rm const.}$ and hence the condition $k = a(t)m_\phi(t)$ continues to be satisfied for a specific wavenumber $k$. 
Thus the (narrow) parametric resonance happens~\cite{Traschen:1990sw,Shtanov:1994ce,Kofman:1994rk,Kofman:1997yn} and the present simple treatment breaks down.

Second, for $w > 1/3$ (or $n>4$), $m_\phi$ redshifts faster than the physical wavenumber $k/a$.\footnote{
	In this case one should take account of the growth of the inflaton fluctuation due to self interactions, which eventually terminates the kination regime and $w$ approaches to $1/3$~\cite{Lozanov:2016hid,Lozanov:2017hjm,Antusch:2021aiw,Garcia:2023dyf,Eroncel:2025bcb}. 
    In this paper we just take the timing of the beginning of the $w=1/3$ regime as a free parameter.
}
In this case, the GW modes that once exit the horizon during inflation can satisfy the condition $k=am_\phi$ and hence are amplified after horizon re-entry (see the left panel of Fig.~\ref{fig:largew}).
Therefore, for these modes, we must take account of \textit{both} the superhorizon enhancement and the particle production from the inflaton oscillation for the same GW mode, though our calculation will show that the latter effect is negligible.
In short, the contribution from inflaton oscillation for $k<k_1$ is smaller than and hidden by the conventional primordial GWs generated during inflation.\footnote{
	This point has been overlooked in Ref.~\cite{Choi:2024ilx}.
}

Third, for $w<1/3$, the resultant GW energy density in the high-frequency limit (\ref{rhoh_ana_high}) at $k=a_{\rm e} m_\phi(t_{\rm e}) \equiv k_2$ is larger than that of the low-frequency limit (\ref{rhoh_ana_low}) at $k = a_{\rm e} H_{\rm e} \equiv k_1$ by a factor $m_\phi(t_{\rm e}) / H_{\rm e}$. Thus we have
\begin{align}
	\frac{\Omega_{\rm GW}(f_2)}{\Omega_{\rm GW}(f_1)} \sim \frac{m_\phi(t_{\rm e})}{H_{\rm e}},
    \label{Ogw_ratio}
\end{align}
which is typically (much) larger than unity (see the right panel of Fig.~\ref{fig:smallw}).
The scaling for $k\ll k_1$ and $k\gg k_2$ are derived above, but the behavior at the intermediate region $k_1 \lesssim k \lesssim k_2$ is difficult to derive in an analytic way. 
In the following sections we numerically evaluate the GW spectrum across these regions for several concrete inflation models.

\subsection{Parameter dependence} \label{sec:temp}

\begin{figure}
\centering
\includegraphics[width=1.0\textwidth]{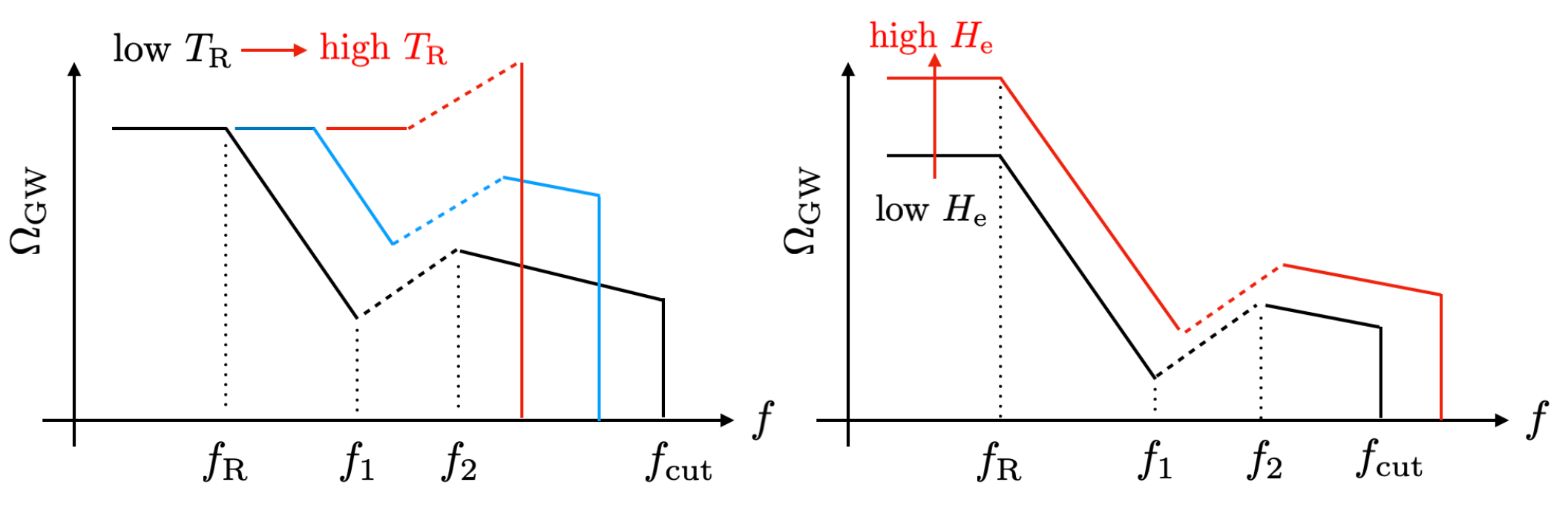}
\caption{Parameter dependence of $\Omega_{\rm GW}$ for $w=0$. 
$T_{\rm R}$ dependence is shown in the left panel, while $H_{\rm e}$ dependence is shown in the right panel.
In the right panel, $H_{\rm e}$ is varied while the ratio $H_{\rm e}/m_\phi$ is fixed.}
\label{fig:TR}
\end{figure}

While the GW spectrum for any parameter choice is deduced from the expressions given in previous section, it will be helpful for readers to more explicitly show how the GW spectrum changes when the parameters are varied.
A typical GW spectrum is shown in Fig.~\ref{fig:TR} for $w=0$. Characteristic frequencies are given by
\begin{align}
	&f_{\rm R} = \frac{a_{\rm R}H_{\rm R}}{2\pi a_0}\simeq 2.7\times 10^2\,{\rm Hz}\left(\frac{T_{\rm R}}{10^{10}\,{\rm GeV}}\right)\left(\frac{g_*(T_{\rm R})}{106.75}\right)^{\frac{1}{6}},\\
    &f_{\rm 1} = \frac{a_{\rm e}H_{\rm e}}{2\pi a_0} \simeq 5.1\times 10^6\,{\rm Hz}\left(\frac{H_{\rm e}}{10^{12}\,{\rm GeV}}\right)^{\frac{1}{3}}\left(\frac{T_{\rm R}}{10^{10}\,{\rm GeV}}\right)^{\frac{1}{3}}\left(\frac{g_*(T_{\rm R})}{g_{*s}(T_{\rm R})}\right),\\
    &f_{\rm 2} = \frac{a_{\rm e}m_\phi}{2\pi a_0} \simeq 5.1\times 10^7\,{\rm Hz}\left(\frac{m_{\phi}}{10^{13}\,{\rm GeV}}\right)\left(\frac{10^{12}\,{\rm GeV}}{H_{\rm e}}\right)^{\frac{2}{3}}\left(\frac{T_{\rm R}}{10^{10}\,{\rm GeV}}\right)^{\frac{1}{3}}\left(\frac{g_*(T_{\rm R})}{g_{*s}(T_{\rm R})}\right)
    ,\\
    &f_{\rm cut} = \frac{a_{\rm R}m_\phi}{2\pi a_0}\simeq 1.9\times 10^{13}\,{\rm Hz}\left(\frac{m_{\phi}}{10^{13}\,{\rm GeV}}\right)\left(\frac{10^{10}\,{\rm GeV}}{T_{\rm R}}\right)\left(\frac{106.75}{g_{*s}(T_{\rm R})}\right)^{\frac{1}{3}}.
\end{align}
$\Omega_{\rm GW}$ at each frequency is given by substituting $k=2\pi a_0 f$ to (\ref{OGW_ana_low}) or (\ref{OGW_ana_high}).
First, let us vary $T_{\rm R}$ while fixing $H_{\rm e}$ and $m_\phi$. By making $T_{\rm R}$ higher, $f_{\rm R}, f_1$ and $f_2$ become larger while $f_{\rm cut}$ becomes smaller. $\Omega_{\rm GW}(f<f_{\rm R})$ is independent of $T_{\rm R}$. In the high $T_{\rm R}$ limit, $f_{\rm R}$ and $f_1$ are merged, and $f_2$ and $f_{\rm cut}$ are merged.
Next let us vary $H_{\rm e}$ while fixing $T_{\rm R}$ and the ratio $H_{\rm e}/m_\phi$.
$f_{\rm R}$ remains constant, while $f_1, f_2, f_{\rm cut}$ become larger. $\Omega_{\rm GW}$ increases as $\propto H_{\rm e}^2$.
These behaviors are shown Fig.~\ref{fig:TR}.
As repeatedly noted, the shape of $\Omega_{\rm GW}$ in the region $f_1<f<f_2$ is nontrivial and numerically calculated in the next section.

\section{Gravitational wave spectrum in concrete models}\label{sec:GW}

In this section we calculate the primordial GW spectrum, especially focusing on the high frequency tail, based on the formalism developed in the previous section.
We do not much care about the consistency of the scalar spectral index and the tensor-to-scalar ratio with cosmological observations~\cite{Planck:2018vyg}, since it is not difficult to (slightly) modify the inflaton potential in order to make the model consistent with observations, while not affecting the dynamics around/after the end of inflation~\cite{Nakayama:2013jka,Nakayama:2013txa}.
In order to calculate $\Omega_{\rm GW}(f)$, we will take $T_{\rm R} = 10^{10}$\,{\rm GeV} as a representative value in this section.
One can convert the results to the case of higher/lower $T_{\rm R}$ by using the formula in Sec.~\ref{sec:rough} and \ref{sec:temp}.

\subsection{Chaotic inflation}\label{sec:chao}

As the simplest example we start from the quadratic inflaton potential~\cite{Linde:1983gd}
\begin{align}
	V = \frac{1}{2}m_\phi^2\phi^2.
\end{align}
The dimensionless power spectrum of the curvature perturbation is given by~\cite{Lyth:2009imm}
\begin{align}
	\mathcal P_\zeta = \frac{N^2 m_\phi^2}{6\pi^2 M_{\rm Pl}^2},
\end{align}
with $N$ being the e-folding number measured from the end of inflation, while the scalar spectral index $n_s$ and tensor-to-scalar ratio $r$ are given by~\cite{Lyth:2009imm}
\begin{align}
	n_s = 1- \frac{2}{N}, ~~~~~~~~~r= \frac{8}{N}.
\end{align}
The inflaton mass is $m_\phi \sim 1.4\times 10^{13}\,{\rm GeV}$ to reproduce the density perturbation of the universe, $\mathcal P_\zeta \simeq 2.2\times 10^{-9}$~\cite{Planck:2018vyg}.

\begin{figure}
\centering
\includegraphics[width=.45\textwidth]{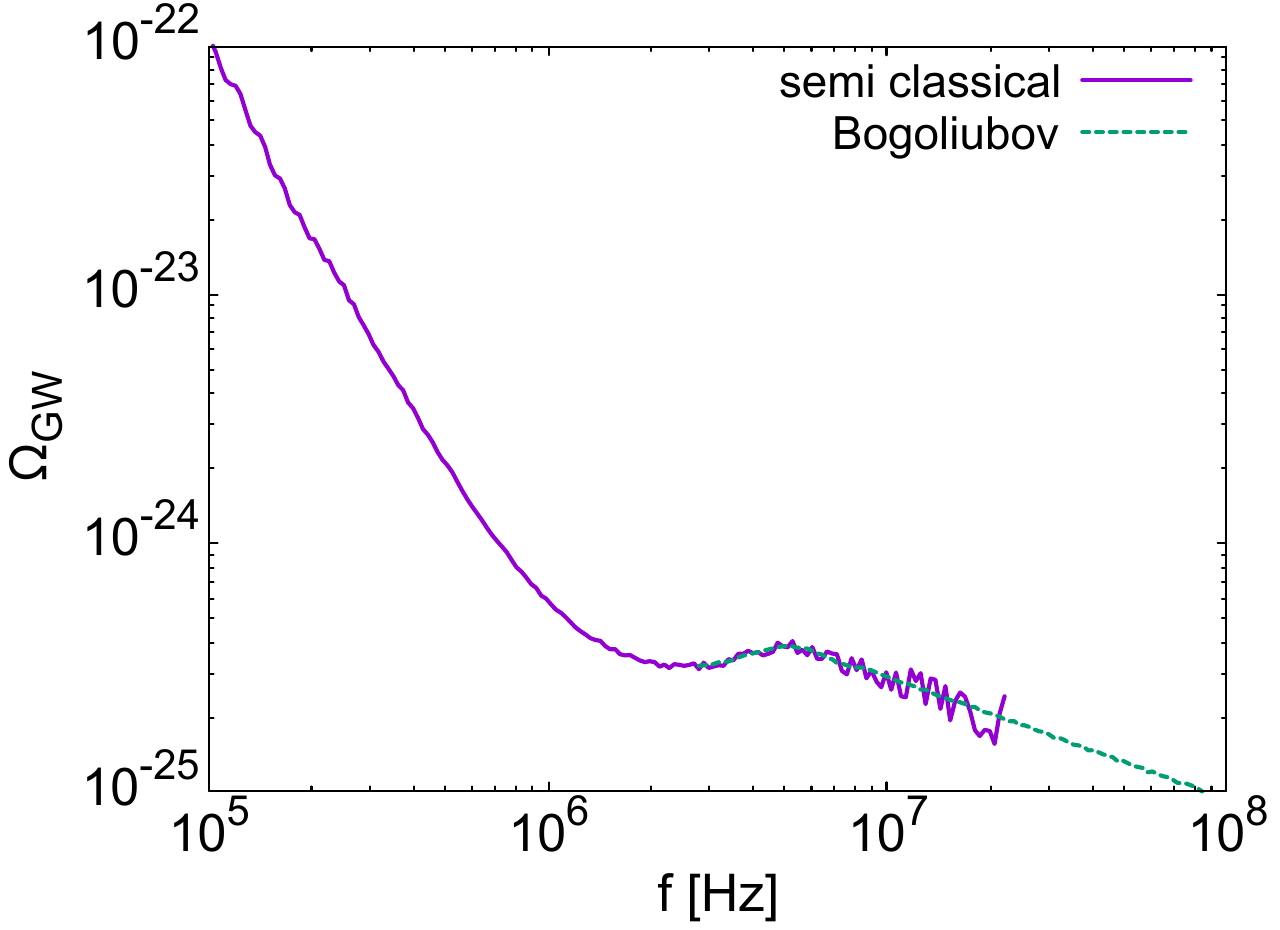}
\includegraphics[width=.45\textwidth]{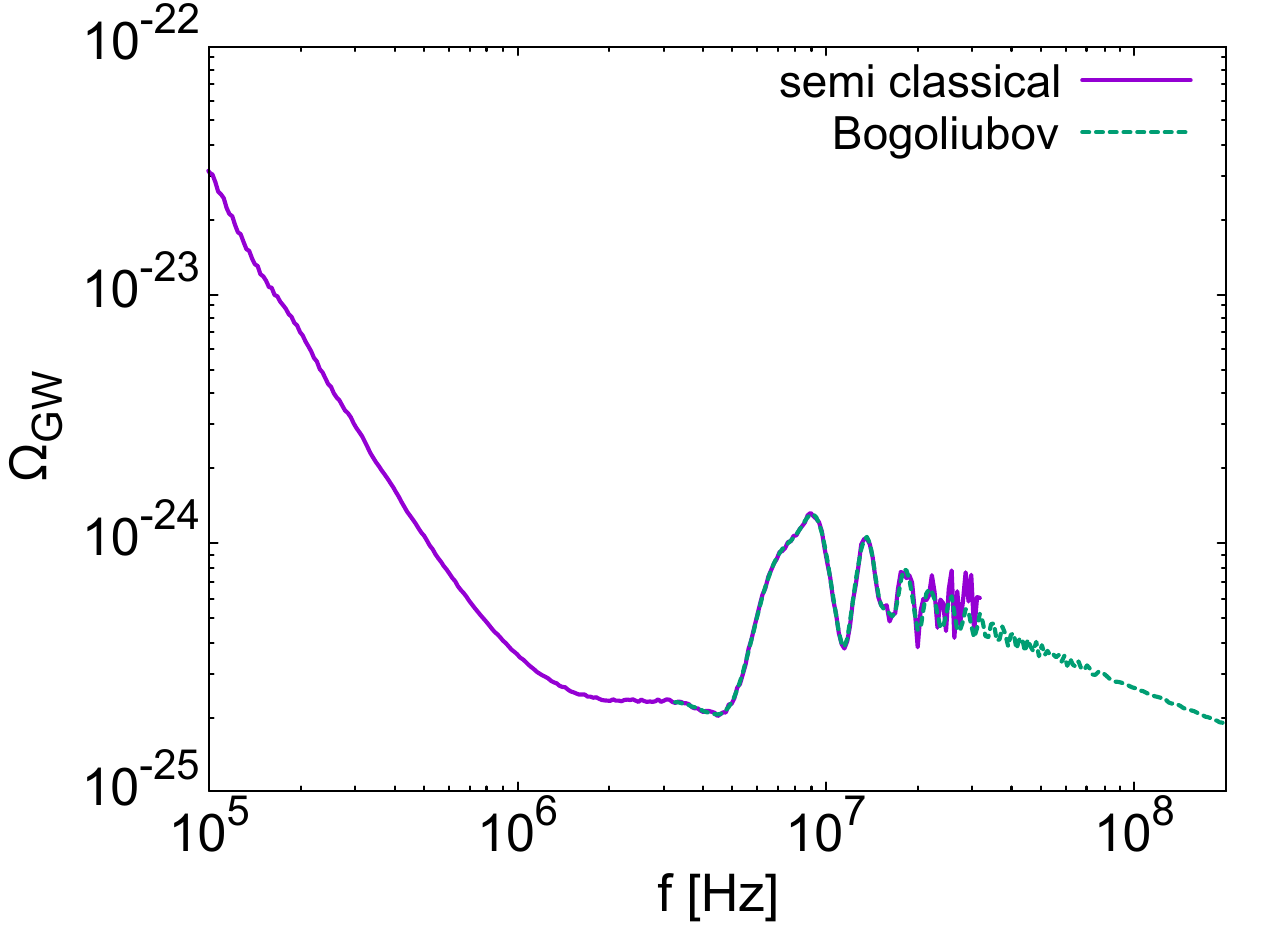}
\caption{
(Left) GW spectrum for quadratic chaotic inflation. (Right) GW spectrum for Starobinsky inflation.
We have taken $T_{\rm R}=10^{10}\,{\rm GeV}$.
Solid line is the result of semi-classical method and the dashed line is the result of Bogoliubov method.
Both give consistent results in the overlapping region where the both methods are applicable.
It is clearly seen that the spectrum scales as $f^{-2}$ in the low frequency limit and $f^{-1/2}$ in the high frequency limit, as expected (see Fig.~\ref{fig:smallw}).
}
\label{fig:GW_chao}
\end{figure}

The result of $\Omega_{\rm GW}(f)$ is shown in the left panel of Fig.~\ref{fig:GW_chao}.
In the low-frequency region we used the semi-classical method, while in the high frequency region we used the Bogoliubov method.
There are overlapping regions where the both methods are useful, and we obtained consistent results.
The behavior $\Omega_{\rm GW}(f) \propto f^{-2}$ for $f<f_1$ and $\Omega_{\rm GW}(f) \propto f^{-1/2}$ for $f>f_2$ are clearly seen, consistent with the  estimation in the previous section, which is schematically shown in Fig.~\ref{fig:smallw} with $w=0$.
In this model $H_{\rm e}$ and $m_\phi$ (hence $f_1$ and $f_2$) are the same order and the existence of the blue region in Fig.~\ref{fig:smallw} is not so obvious.
But still there seems to be a transition region where the spectrum shows a nontrivial behavior.\footnote{
	This peculiar structure had already been seen in Ref.~\cite{Ema:2020ggo}.
}

\subsection{Starobinsky inflation}\label{sec:sta}

In the Starobinsky inflation model~\cite{Starobinsky:1980te}, the $R^2$ term is introduced in the action with $R$ being the Ricci curvature. 
After appropriate conformal transformation to go to the Einstein frame, a scalaron degree of freedom appears, which is identified as the inflaton.
The scalar potential for the inflaton is given by
\begin{align}
	V = \frac{3m_\phi^2 M_{\rm Pl}^2}{4} \left[ 1-\exp\left(-\sqrt{\frac{2}{3}}\frac{\phi}{M_{\rm Pl}} \right) \right]^2.
\end{align}
The dimensionless power spectrum of the curvature perturbation is given by
\begin{align}
	\mathcal P_\zeta = \frac{N^2 m_\phi^2}{24\pi^2 M_{\rm Pl}^2}.
\end{align}
The inflaton mass should be $m_\phi \simeq 3\times 10^{13}\,{\rm GeV}$ for reproducing the cosmological observations.
The scalar spectral index and tensor-to-scalar ratio are given by
\begin{align}
	n_s = 1 - \frac{2}{N},~~~~~~r = \frac{12}{N^2}.
\end{align}
In the minimal Starobinsky model, the reheating temperature is determined by the inflaton decay rate into the Higgs particles as (see e.g. Refs.~\cite{Gorbunov:2010bn,Li:2021fao})
\begin{align}
	T_{\rm R} \simeq 5\times 10^9\,{\rm GeV} \times |1-6\xi|,
\end{align}
where $\xi$ denotes the nonminimal coupling between the Higgs and Ricci curvature.
Thus the choice of $T_{\rm R}=10^{10}\,{\rm GeV}$ is natural in this model.

The result of $\Omega_{\rm GW}(f)$ is shown in the right panel of Fig.~\ref{fig:GW_chao}.
Similar to the chaotic inflation, both the $f^{-2}$ behavior  in the low frequency limit and $f^{-1/2}$ behavior in the high frequency limit are clearly seen.
An intermediate region is a bit broader and the structure is more prominent than the case of chaotic inflation.

\subsection{New inflation}\label{sec:new}

Next we consider new inflation model~\cite{Linde:1981mu,Albrecht:1982wi}, in which the inflaton rolls down the potential toward the symmetry breaking vacuum expectation value.
In this paper we define the new inflation model by the following potential~\cite{Kumekawa:1994gx,Izawa:1996dv,Asaka:1999jb,Senoguz:2004ky,Kohri:2007gq,Nakayama:2012dw,Ema:2017rkk}
\begin{align}
	V = \Lambda^4 \left[ 1 - \left(\frac{\phi}{v}\right)^n \right]^2.
\end{align}
In this model there are two parameters, $\Lambda$ and $v$, and hence there is a degree of freedom to choose the inflationary Hubble scale and the inflaton mass.
The dimensionless power spectrum of the curvature perturbation is given by
\begin{align}
	\mathcal P_\zeta = \frac{1}{12\pi^2}\left[2n \left((n-2) N\right)^{n-1}\right]^{\frac{2}{n-2}} \frac{\Lambda^4}{(v^n M_{\rm Pl}^{n-4})^{\frac{2}{n-2}}}.
\end{align}
Scalar spectral index and tensor-to-scalar ratio is
\begin{align}
	n_s = 1- \frac{2}{N}\frac{n-1}{n-2},~~~~~~r =\frac{16n}{N(n-2)} \left[\frac{1}{2n(n-2)N}\frac{v^2}{M_{\rm Pl}^2}\right]^{\frac{n}{n-2}}.
\end{align}
The inflaton mass around the minimum is
\begin{align}
	m_\phi = \frac{\sqrt{2}n \Lambda^2}{v} = \frac{\sqrt{6}n M_{\rm Pl}}{v} H_{\rm e}~~~~~~\leftrightarrow~~~~~~
    \frac{H_{\rm e}}{m_\phi} =  \frac{v}{\sqrt{6}n M_{\rm Pl}}.
\end{align}
Therefore there is a large hierarchy between the inflaton mass and inflationary Hubble scale for smaller $v$, and hence $f_2 / f_1$ also becomes large (see Fig.~\ref{fig:smallw}).

As an example, let us consider the case of $n=4$. In this case, to reproduce the density perturbation $\mathcal P_\zeta=2.2\times 10^{-9}$, we obtain
\begin{align}
	\Lambda \simeq 9\times 10^{14}\,{\rm GeV} \left(\frac{v}{M_{\rm Pl}}\right) \left(\frac{60}{N}\right)^{\frac{3}{4}}.
\end{align}
The inflaton mass and inflationary Hubble scale are given by
\begin{align}
	m_\phi \simeq 2\times 10^{12}\,{\rm GeV}  \left(\frac{v}{M_{\rm Pl}}\right)\left(\frac{60}{N}\right)^{\frac{3}{2}},~~~~~~
	H_{\rm e} \simeq  2\times 10^{11}\,{\rm GeV} \left(\frac{v}{M_{\rm Pl}}\right)^2  \left(\frac{60}{N}\right)^{\frac{3}{2}}.
\end{align}
\begin{figure}
\centering
\includegraphics[width=.45\textwidth]{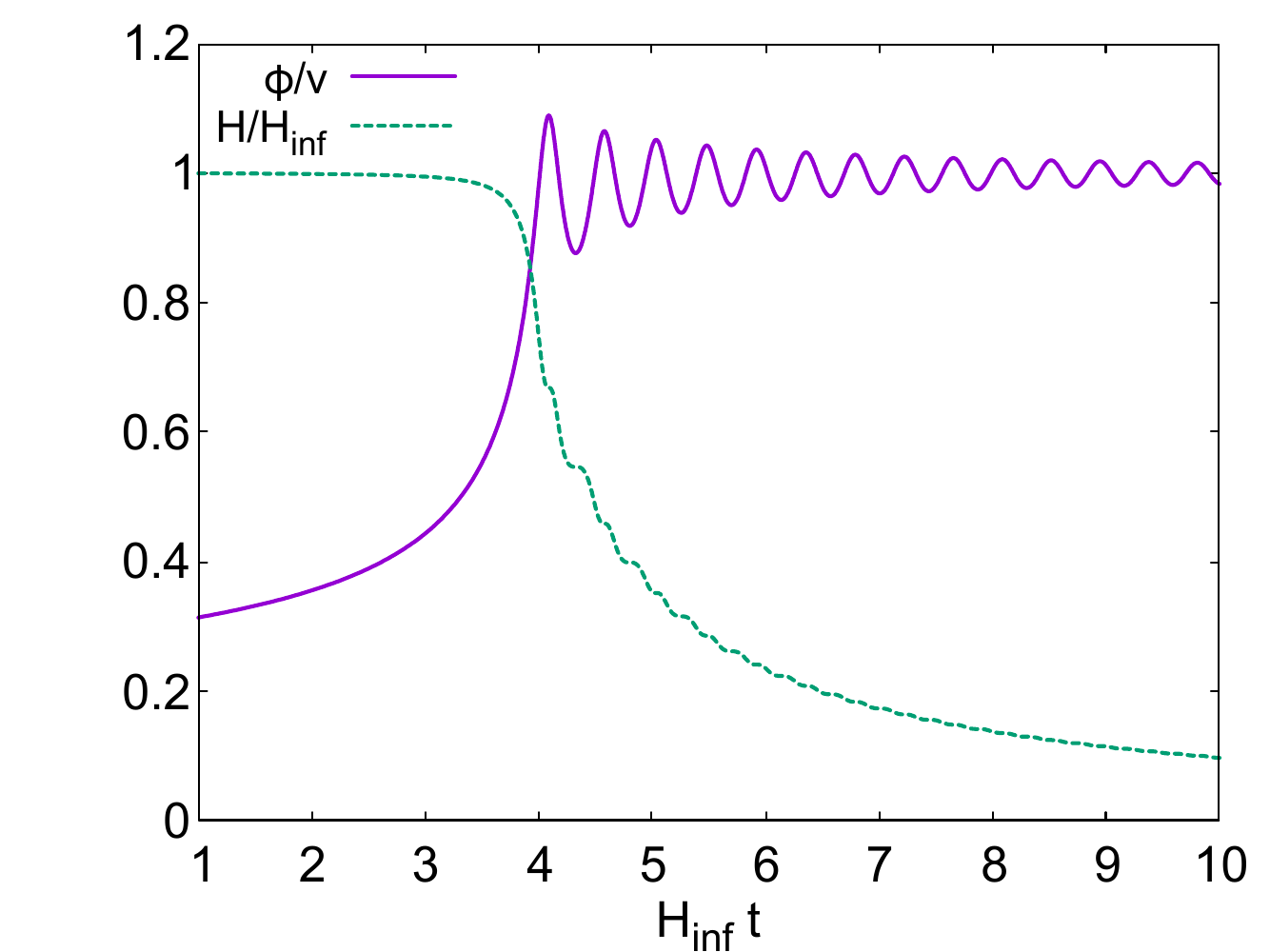}
\includegraphics[width=.45\textwidth]{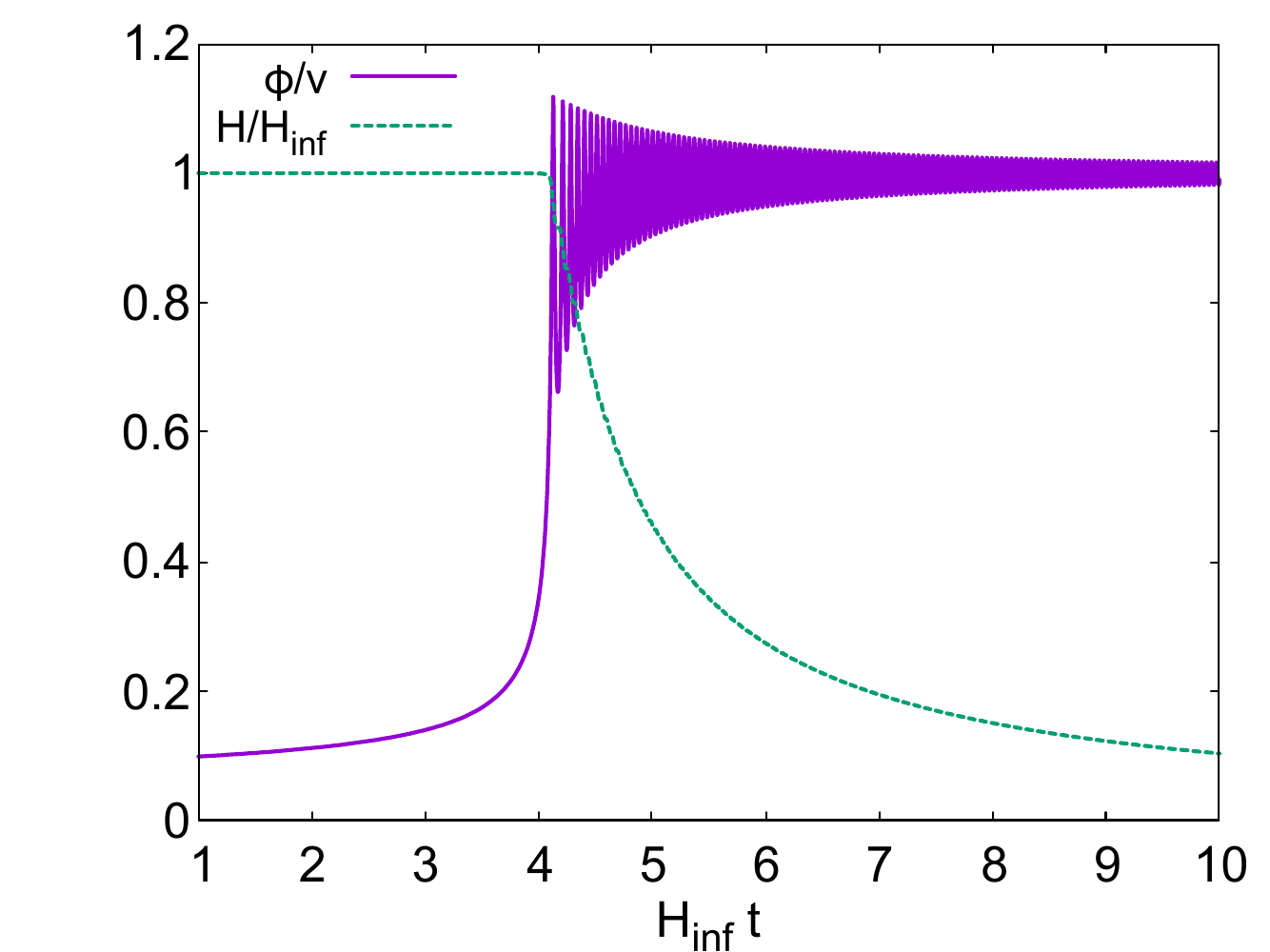}
\caption{Time evolution of the homogeneous inflaton field $\phi/v$ (solid) and the Hubble parameter $H/H_{\rm inf}$ (dashed) in the new inflation model. We have taken $v/M_{\rm Pl} = 1$ in the left panel and $v/M_{\rm Pl} = 0.1$ in the right panel.}
\label{fig:newinf}
\end{figure}

Fig.~\ref{fig:newinf} shows the time evolution of the inflaton field for $v/M_{\rm Pl}=1$ (left panel) and $v/M_{\rm Pl}=0.1$ (right panel).
It is seen that the oscillation time scale, $m_\phi^{-1}$, which is much shorter than the Hubble scale $H_{\rm e}^{-1}$, soon appears after inflation ends.
Thus the transition from inflationary era to the inflaton oscillation era is close to instantaneous and a rough picture depicted in Fig.~\ref{fig:smallw} is expected to apply.\footnote{
	In reality, the inflaton fluctuations may grow as the inflaton starts to oscillate~\cite{Brax:2010ai,Antusch:2015nla,Antusch:2015vna}. This effect is stronger for smaller $v$ and we can neglect such an effect for $v/M_{\rm Pl} \gtrsim 0.1$, as shown in Ref.~\cite{Antusch:2015nla}. In this paper we only consider the case of $v/M_{\rm Pl}\gtrsim 0.1$ and do not treat the case of smaller $v$.}

\begin{figure}
\centering
\includegraphics[width=.45\textwidth]{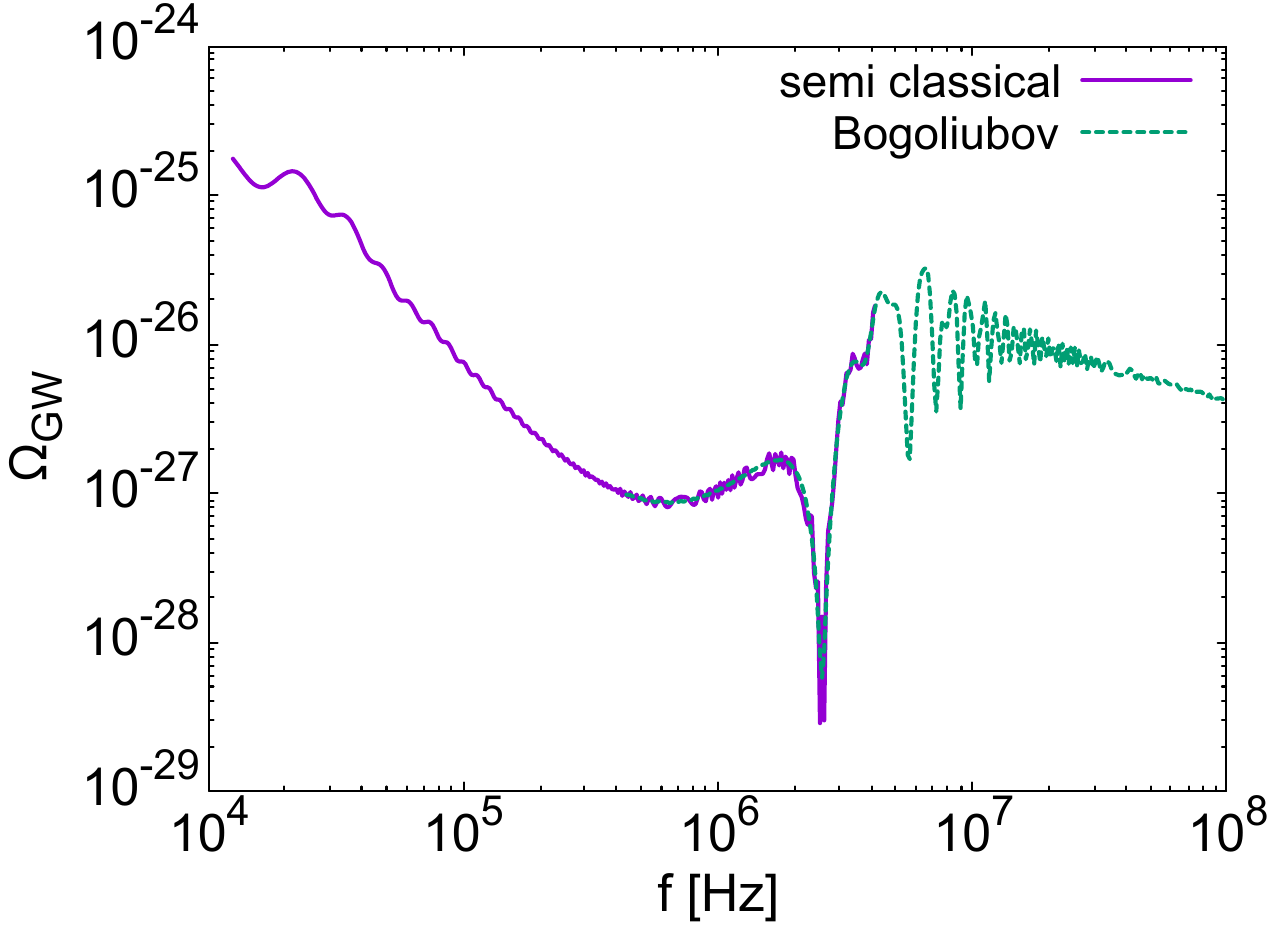}
\includegraphics[width=.45\textwidth]{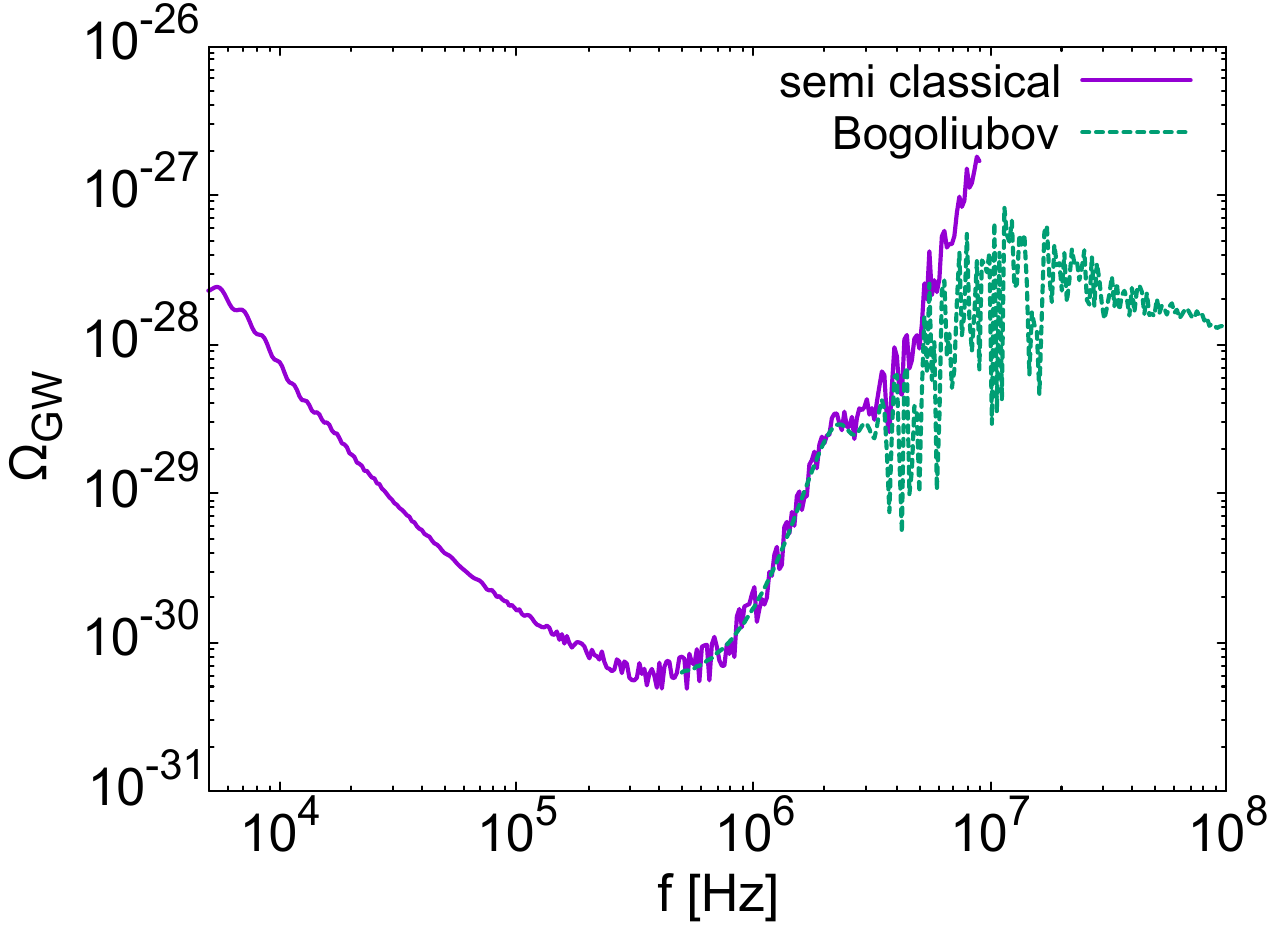}
\caption{GW spectrum in new inflation model. We have taken $v/M_{\rm Pl} = 1$ in the left panel and $v/M_{\rm Pl} = 0.1$ in the right panel.}
\label{fig:GW_new}
\end{figure}

The result of numerical calculation for the GW spectrum are shown in Fig.~\ref{fig:GW_new}.
We have taken $v/M_{\rm Pl} = 1$ (left panel) and $v/M_{\rm Pl}=0.1$ (right panel).
We can see the $f^{-2}$ and $f^{-1/2}$ behavior in the low-frequency ($f< f_1$) and high-frequency ($f>f_2$) limit, respectively.
They are perfectly consistent with the rough estimation in Sec.~\ref{sec:rough} with $w=0$ and also a schematic picture shown in Fig.~\ref{fig:smallw}.
Our numerical simulation revealed the structure in the intermediate frequency region $f_1\lesssim f \lesssim f_2$.
Although there are some nontrivial oscillatory behavior there, as an overall structure, the spectrum roughly monotonically increases from $\Omega_{\rm GW}(f_1)$ to $\Omega_{\rm GW}(f_2)$.


\subsection{Attractor inflation}\label{sec:att}

Finally let us consider the $\alpha$-attractor T model~\cite{Kallosh:2013hoa,Kallosh:2013yoa,Galante:2014ifa}. This model utilizes the kinetic term like $\mathcal L\sim -f(\phi)(\partial\phi)^2/2$ with $f(\phi) \sim (1-\phi^2/\Lambda^2)^{-2}$, which is divergent at $\phi = \Lambda$. 
After redefining the inflaton field so that it is canonically normalized (and rewrite the canonical inflaton as $\phi$), the scalar potential typically looks like
\begin{align}
	V = \frac{\lambda}{n} \left[ \Lambda \tanh\left(\frac{|\phi|}{\Lambda}\right)\right]^n,
\end{align}
where $\lambda$ is a constant. The dimensionless power spectrum is given by
\begin{align}
	\mathcal P_\zeta = \frac{N^2 \lambda \Lambda^{n-2}}{3n \pi^2 M_{\rm Pl}^2}.
\end{align}
The scalar spectral index and tensor-to-scalar ratio are given by
\begin{align}
	n_s = 1-\frac{2}{N},~~~~~~~~~r = \frac{2}{N^2} \left(\frac{\Lambda}{M_{\rm Pl}}\right)^2.
\end{align}

For $n=2$, the inflaton mass is identified as $m_\phi = \sqrt{\lambda}$ and it is determined to reproduced the observed density perturbation.
The inflaton mass and the Hubble scale are given by
\begin{align}
	m_\phi \simeq 1.4\times 10^{13}\,{\rm GeV}\,\left(\frac{60}{N}\right),~~~~~~\frac{H_{\rm e}}{m_\phi} = \frac{\Lambda}{\sqrt 6 M_{\rm Pl}}.
\end{align}
Therefore, for $\Lambda \ll M_{\rm Pl}$ we have hierarchy $H_{\rm e}\ll m_\phi$.
For general $n$, the coupling constant $\lambda$ is given by
\begin{align}
	\lambda M_{\rm Pl}^{n-4} = \frac{3\pi^2 n \mathcal P_\zeta}{N^2}\left(\frac{M_{\rm Pl}}{\Lambda}\right)^{n-2},
\end{align}
and defining $m_\phi^2(t_{\rm e})=\lambda \Lambda^{n-2}$, we obtain
\begin{align}
	m_\phi(t_{\rm e}) \simeq \frac{\sqrt{3\pi^2 n \mathcal P_\zeta}}{N} M_{\rm Pl},~~~~~~
    \frac{H_{\rm e}}{m_\phi(t_{\rm e})} \simeq \frac{\Lambda}{\sqrt{3n}M_{\rm Pl}}.
\end{align}
As noted in Sec.~\ref{sec:rough}, for $n>2$ the initially homogeneous inflaton field eventually becomes highly inhomogeneous due to the self-resonance effect~\cite{Lozanov:2016hid,Lozanov:2017hjm,Antusch:2021aiw,Garcia:2023dyf,Eroncel:2025bcb}.
However, it takes e-foldings of $\Delta N_e \sim \ln\left(10^4 \Lambda/M_{\rm Pl}\right)$~\cite{Lozanov:2017hjm,Garcia:2023dyf} from the inflation end until the fragmentation effect becomes important. For our parameter choices below, this is long enough and we can safely neglect such an effect.

\begin{figure}
\centering
\includegraphics[width=.45\textwidth]{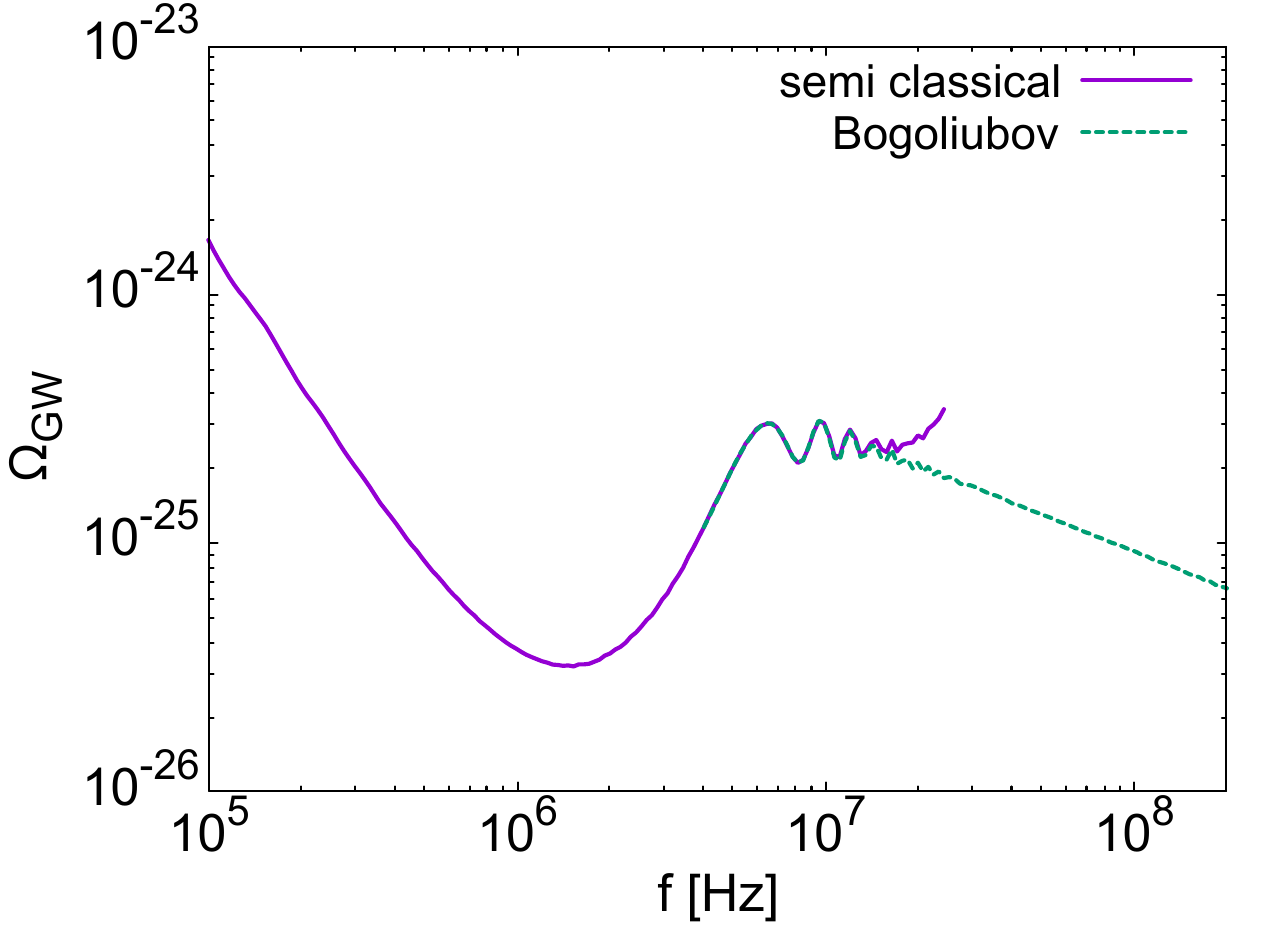}
\includegraphics[width=.45\textwidth]{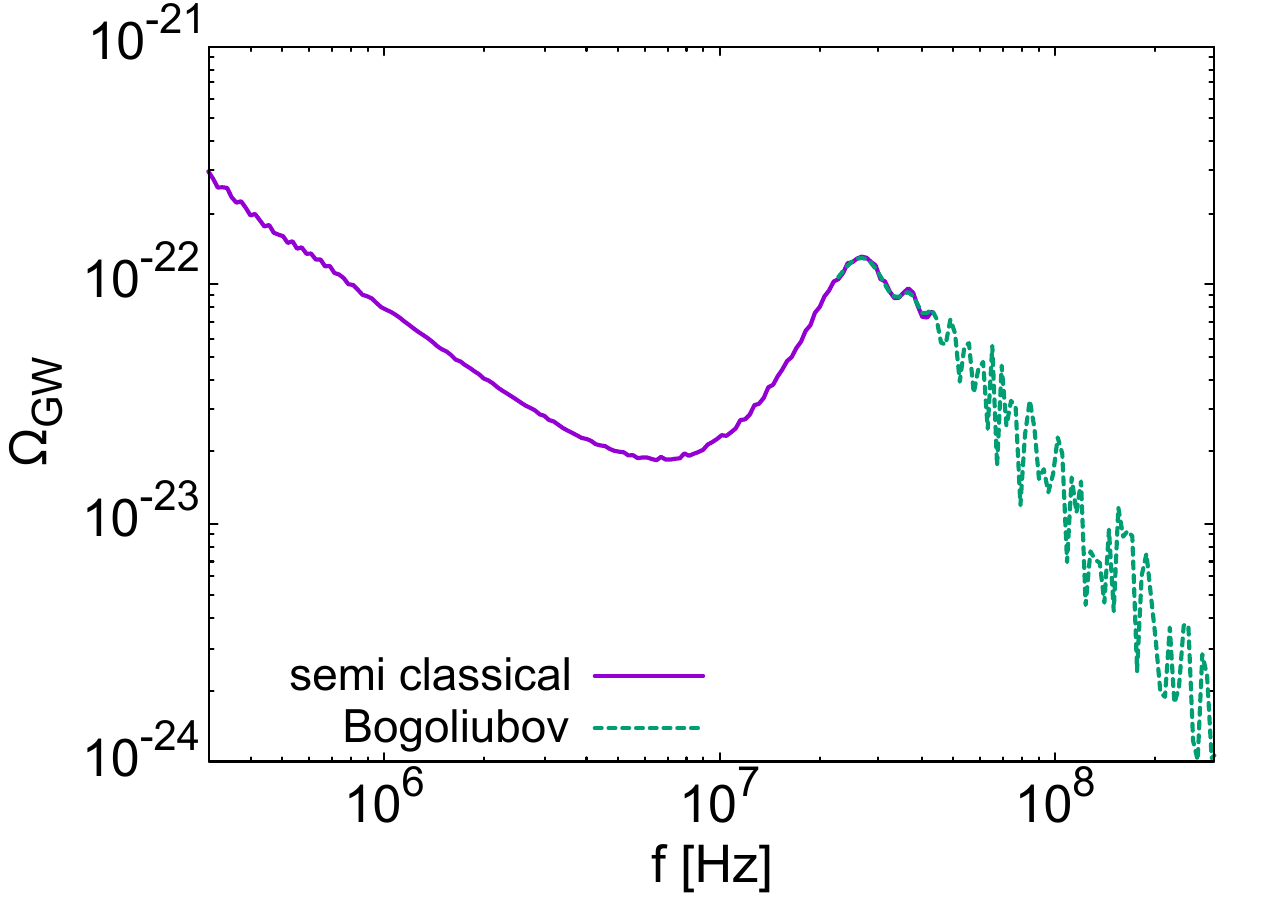}
\includegraphics[width=.45\textwidth]{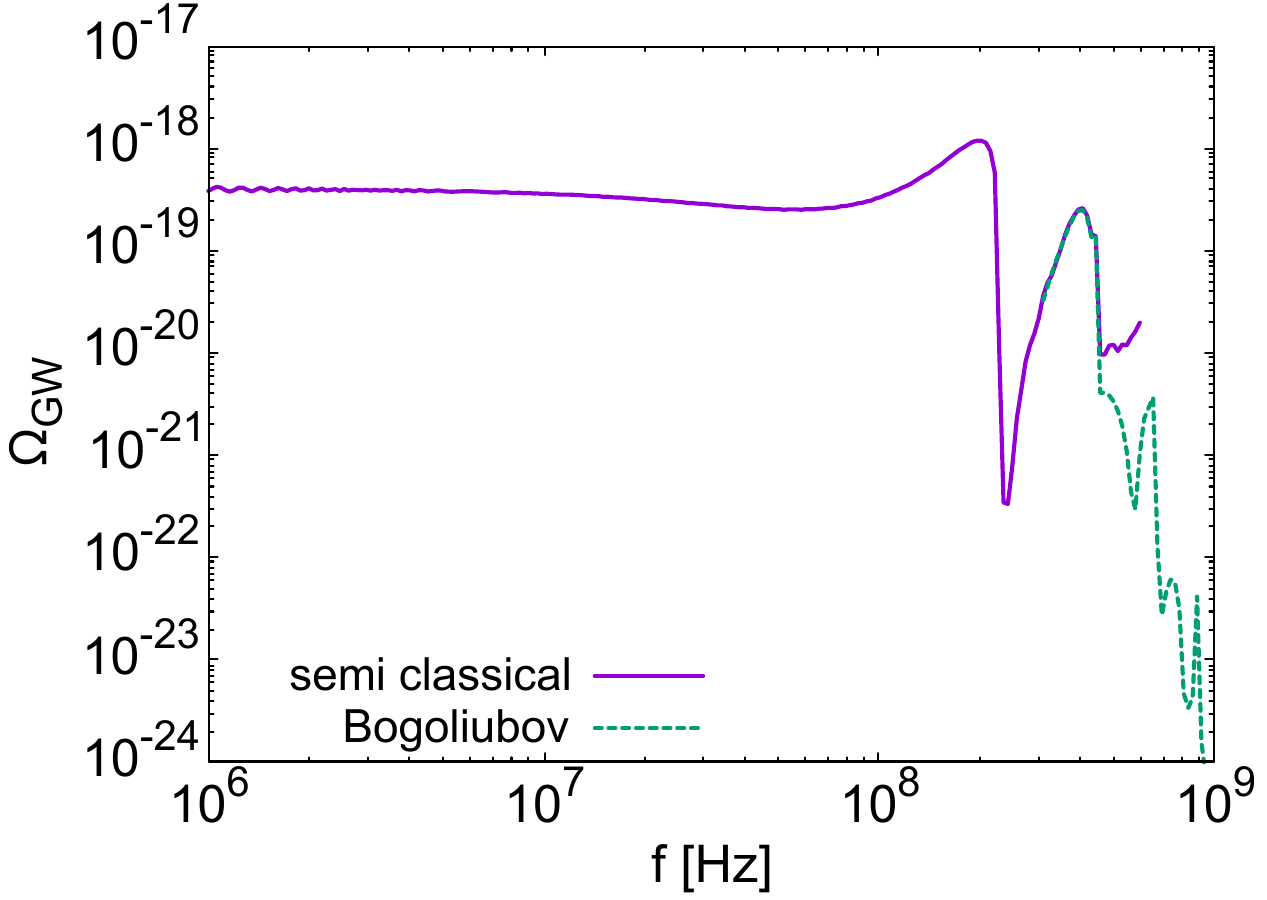}
\includegraphics[width=.45\textwidth]{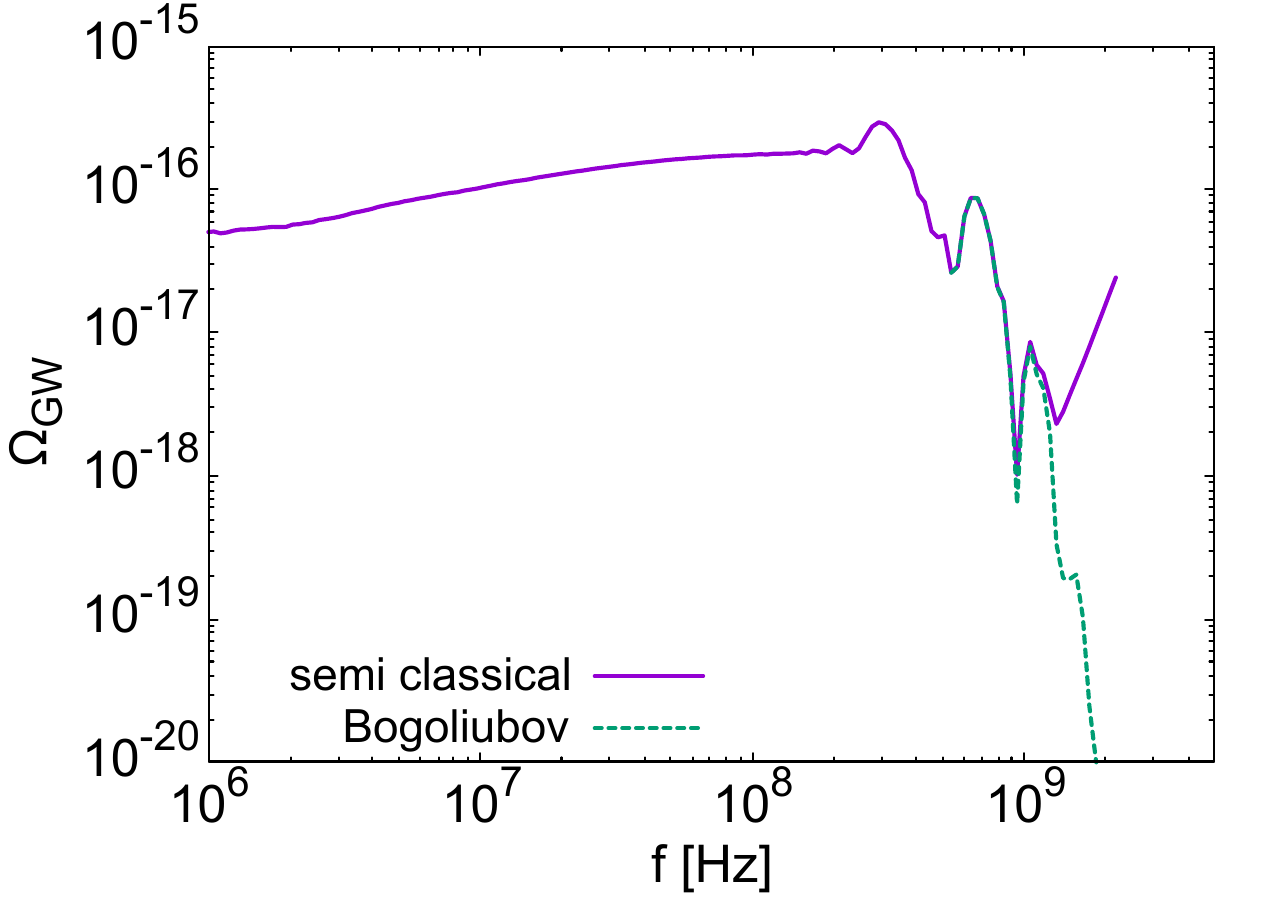}
\caption{GW spectrum in the T-model with $n=2$ (upper left), $n=2.5$ (upper right), $n=4$ (lower left) and $n=6$ (lower right). We have taken $\Lambda/M_{\rm Pl} = 0.5$.}
\label{fig:GW_Tmodel}
\end{figure}

The result of numerical calculation for the GW spectrum are shown in Fig.~\ref{fig:GW_Tmodel} for $n=2$ (upper left), $n=2.5$ (upper right), $n=4$ (lower left) and $n=6$ (lower right). We have taken $\Lambda/M_{\rm Pl} = 0.5$.
The behavior for $n=2$ is more or less similar to the previous cases like Starobinsky or new inflation models since we have $w=0$: we can see both $f^{-2}$ and $f^{-1/2}$ scaling at low- and high-frequency limit, respectively.
Similarly, for $n=2.5$, we have $f^{-1}$ and $f^{-2}$ scalings, which are also consistent with the estimate in Sec.~\ref{sec:rough}.
For $n=4$, we effectively have $w=1/3$ during the inflaton oscillation and hence we have flat GW spectrum for those experienced the superhorizon regime during inflation, as expected from analytic estimate Eq.~(\ref{rhoh_ana_low}).
GWs with higher frequency never experience significant particle production from the inflaton oscillation because they never hits the condition $k= a(t)m_\phi(t)$ (see left panel of Fig.~\ref{fig:largew} with $w=1/3$).
Thus the spectrum rapidly decreases for $f\gg f_2$.

\section{Conclusions and discussion}
\label{sec:conc}

Inflation produces stochastic GW background with an extremely wide frequency range.
Even very high frequency GWs which never experience the super-Hubble-radius regime are amplified through the post-inflationary dynamics of the inflaton.
It is not very difficult to analytically estimate the resulting GW spectrum both in the low and high frequency limit, but there is an intermediate frequency region in which an analytic estimation is difficult.
In particular, in a standard reheating model in which the inflaton oscillates with a quadratic potential, there is a gap in the GW spectrum between the low- and high-frequency regimes.
This gap is represented by a ratio between the inflaton mass and Hubble scale (\ref{Ogw_ratio}) (see also Fig.~\ref{fig:smallw}), which is typically (much) larger than unity, and the spectrum that connects between two values is nontrivial. 
We have developed useful methods to precisely calculate the GW spectrum including such an intermediate frequency range, and performed numerical calculations for some concrete inflation models.
We confirmed the analytic results for the low and high frequency limit, while we found a peculiar structure in the GW spectrum in the intermediate range. 
Such structures may contain detailed information about the inflaton potential and be useful for distinguishing the inflation model.

Let us mention applicability of our results.
In deriving GW spectra in various models, we assumed that the inflaton is spatially homogeneous. As mentioned several times in the main text, the spatial inhomogeneity of the inflaton may develop due to its self-resonance and may fragment during the reheating.
We have not taken into account such an effect in this study. At least for parameters we used in our numerical analysis, the effect of fragmentation is likely to be negligible since it takes rather long time until the perturbation develops sufficiently~\cite{Lozanov:2017hjm,Garcia:2023dyf}.
However, for models with lower inflation scale, such an effect may not be negligible. Further studies are required for more precise estimation of GWs produced during the reheating.

While such high frequency GW spectrum shows interesting features, it should be noticed that its abundance is too small to detect in near future.
Moreover, it may be hidden by the other GW sources: for example, bremsstrahlung GWs emitted at the inflaton decay~\cite{Nakayama:2018ptw,Barman:2023ymn,Barman:2023rpg,Hu:2024awd,Hu:2024bha}, a possible inflaton decay into the graviton pair~\cite{Ema:2021fdz,Mudrunka:2023wxy,Tokareva:2023mrt,Strumia:2025dfn,Nakayama:2025xkn}, back-scattering of high-energy particles off the inflaton~\cite{Xu:2024fjl,Bernal:2025lxp,Xu:2025wjq}, GWs from scattering of Standard Model particles~\cite{Ghiglieri:2015nfa,Ghiglieri:2020mhm,Ringwald:2020ist,Ghiglieri:2022rfp}, and GWs from large density fluctuations through preheating~\cite{Khlebnikov:1997di,Easther:2006vd,Easther:2006gt,Garcia-Bellido:2007nns,Garcia-Bellido:2007fiu,Dufaux:2007pt,Dufaux:2008dn}.
However, the frequency dependence of these sources are much different from the one we derived and it is possible that the GW spectrum derived in this paper gives dominant contribution in some frequency range and hence still valuable to give precise prediction of the primordial GW spectrum.

\section*{Acknowledgment}

This work was supported by World Premier International Research Center Initiative (WPI), MEXT, Japan.
This work was also supported by JSPS KAKENHI (Grant Number 24K07010 [KN]).

\appendix

\section{Convention} \label{sec:conv}

Let us define the tensor perturbation $h_{ij}$ as
\begin{align}
	ds^2=-dt^2 + a^2(t)(\delta_{ij} + c_1 h_{ij}(t,\vec x)) dx^idx^j,
\end{align}
with a numerical factor $c_1$, which is usually taken to be $1$ or $2$ depending on the literature.
When the GW propagates along the $z$ direction, it is further decomposed as
\begin{align}
	h_{ij} = c_2 \begin{pmatrix}
		h_+ & h_\times & 0 \\
		h_\times & -h_+ & 0\\
		0 & 0 & 0
	 \end{pmatrix},
\end{align}
where $c_2$ is another numerical factor, which is usually taken to be $1$ or $1/\sqrt 2$ depending on the literature, corresponding to the convention of the polarization sum $e_{ij}^\lambda e_{ij}^{*\lambda'} =  2c_2^2\delta_{\lambda\lambda'}$, when $h_{ij}$ is expanded as Eq.~(\ref{hij_expand}).
The Einstein-Hilbert action is expanded as
\begin{align}
	\mathcal L &= \frac{M_{\rm Pl}^2}{2} R 
	= -\frac{c_1^2 M_{\rm Pl}^2}{8} (\partial_\mu h_{\rho\sigma})^2
	= -\frac{c_1^2c_2^2M_{\rm Pl}^2}{4} \sum_{\lambda=+,\times} (\partial_\mu h_\lambda)^2,
\end{align}
in the transverse-traceless gauge $\partial_i h_{ij} = h^i_i = 0$ for the on-shell graviton.
The canonical graviton is therefore
\begin{align}
	h_\lambda^{\rm (can)} \equiv \frac{c_1 c_2 M_{\rm Pl}}{\sqrt 2}h_\lambda.
\end{align}

With these definitions, the dimensionless tensor power spectrum in the de-Sitter universe in the superhorizon limit is
\begin{align}
	\Delta^2_{h^{\rm (can)}} = c_3\,\left(\frac{H_{\rm inf}}{2\pi}\right)^2~~~~~~\to~~~~~~
	\Delta^2_{h} = \frac{c_3}{c_1^2 c_2^2\,2\pi^2}\left(\frac{H_{\rm inf}}{M_{\rm Pl}}\right)^2,
\end{align}
where $c_3$ is a numerical factor. If one defines $\Delta^2_{h^{\rm (can)}}$ as a sum of two polarizations, we should choose $c_3=2$.
If one defines $\Delta^2_{h^{\rm (can)}}$ for each polarization, we may take $c_3=1$.
The dimensionless power spectrum of the curvature perturbation is given by
\begin{align}
	\Delta_\zeta^2 =  \left( \frac{H_{\rm inf}}{\dot \phi} \frac{H_{\rm inf}}{2\pi}\right)^2 = \frac{1}{2\epsilon M_{\rm Pl}^2} \left(\frac{H_{\rm inf}}{2\pi}\right)^2.
\end{align}
where $\epsilon$ is a slow-roll parameter. The tensor-to-scalar ratio is then defined by
\begin{align}
	r \equiv c_4 \frac{\Delta^2_{h}}{\Delta_\zeta^2} = \frac{c_3 c_4}{c_1^2 c_2^2}\times 4\epsilon,
\end{align}
where $c_4$ is another numerical factor. 
In this paper we take $c_1=1, c_2=1/\sqrt{2}, c_3=2, c_4=1$,\footnote{
    This convention is the same as e.g. Ref.~\cite{Baumann:2022mni}.
} which yields $r=16\epsilon$.

\bibliographystyle{utphys}
\bibliography{ref}

\providecommand{\href}[2]{#2}\begingroup\raggedright\begin{thebibliography}{10}

\bibitem{Starobinsky:1980te}
A.~A. Starobinsky, ``{A New Type of Isotropic Cosmological Models Without
  Singularity},'' \href{http://dx.doi.org/10.1016/0370-2693(80)90670-X}{{\em
  Phys. Lett. B} {\bfseries 91} (1980) 99--102}.

\bibitem{Guth:1980zm}
A.~H. Guth, ``{The Inflationary Universe: A Possible Solution to the Horizon
  and Flatness Problems},''
  \href{http://dx.doi.org/10.1103/PhysRevD.23.347}{{\em Phys. Rev. D}
  {\bfseries 23} (1981) 347--356}.

\bibitem{Sato:1981qmu}
K.~Sato, ``{First-order phase transition of a vacuum and the expansion of the
  Universe},'' \href{http://dx.doi.org/10.1093/mnras/195.3.467}{{\em Mon. Not.
  Roy. Astron. Soc.} {\bfseries 195} no.~3, (1981) 467--479}.

\bibitem{Kazanas:1980tx}
D.~Kazanas, ``{Dynamics of the Universe and Spontaneous Symmetry Breaking},''
  \href{http://dx.doi.org/10.1086/183361}{{\em Astrophys. J. Lett.} {\bfseries
  241} (1980) L59--L63}.

\bibitem{Linde:1981mu}
A.~D. Linde, ``{A New Inflationary Universe Scenario: A Possible Solution of
  the Horizon, Flatness, Homogeneity, Isotropy and Primordial Monopole
  Problems},'' \href{http://dx.doi.org/10.1016/0370-2693(82)91219-9}{{\em Phys.
  Lett. B} {\bfseries 108} (1982) 389--393}.

\bibitem{Albrecht:1982wi}
A.~Albrecht and P.~J. Steinhardt, ``{Cosmology for Grand Unified Theories with
  Radiatively Induced Symmetry Breaking},''
  \href{http://dx.doi.org/10.1103/PhysRevLett.48.1220}{{\em Phys. Rev. Lett.}
  {\bfseries 48} (1982) 1220--1223}.

\bibitem{Starobinsky:1979ty}
A.~A. Starobinsky, ``{Spectrum of relict gravitational radiation and the early
  state of the universe},'' {\em JETP Lett.} {\bfseries 30} (1979) 682--685.

\bibitem{Allen:1987bk}
B.~Allen, ``{The Stochastic Gravity Wave Background in Inflationary Universe
  Models},'' \href{http://dx.doi.org/10.1103/PhysRevD.37.2078}{{\em Phys. Rev.
  D} {\bfseries 37} (1988) 2078}.

\bibitem{Turner:1990rc}
M.~S. Turner and F.~Wilczek, ``{Relic gravitational waves and extended
  inflation},'' \href{http://dx.doi.org/10.1103/PhysRevLett.65.3080}{{\em Phys.
  Rev. Lett.} {\bfseries 65} (1990) 3080--3083}.

\bibitem{Turner:1993vb}
M.~S. Turner, M.~J. White, and J.~E. Lidsey, ``{Tensor perturbations in
  inflationary models as a probe of cosmology},''
  \href{http://dx.doi.org/10.1103/PhysRevD.48.4613}{{\em Phys. Rev. D}
  {\bfseries 48} (1993) 4613--4622},
  \href{http://arxiv.org/abs/astro-ph/9306029}{{\ttfamily
  arXiv:astro-ph/9306029}}.

\bibitem{Turner:1996ck}
M.~S. Turner, ``{Detectability of inflation produced gravitational waves},''
  \href{http://dx.doi.org/10.1103/PhysRevD.55.R435}{{\em Phys. Rev. D}
  {\bfseries 55} (1997) R435--R439},
  \href{http://arxiv.org/abs/astro-ph/9607066}{{\ttfamily
  arXiv:astro-ph/9607066}}.

\bibitem{Smith:2005mm}
T.~L. Smith, M.~Kamionkowski, and A.~Cooray, ``{Direct detection of the
  inflationary gravitational wave background},''
  \href{http://dx.doi.org/10.1103/PhysRevD.73.023504}{{\em Phys. Rev. D}
  {\bfseries 73} (2006) 023504},
  \href{http://arxiv.org/abs/astro-ph/0506422}{{\ttfamily
  arXiv:astro-ph/0506422}}.

\bibitem{Boyle:2005se}
L.~A. Boyle and P.~J. Steinhardt, ``{Probing the early universe with
  inflationary gravitational waves},''
  \href{http://dx.doi.org/10.1103/PhysRevD.77.063504}{{\em Phys. Rev. D}
  {\bfseries 77} (2008) 063504},
  \href{http://arxiv.org/abs/astro-ph/0512014}{{\ttfamily
  arXiv:astro-ph/0512014}}.

\bibitem{Seto:2003kc}
N.~Seto and J.~Yokoyama, ``{Probing the equation of state of the early universe
  with a space laser interferometer},''
  \href{http://dx.doi.org/10.1143/JPSJ.72.3082}{{\em J. Phys. Soc. Jap.}
  {\bfseries 72} (2003) 3082--3086},
  \href{http://arxiv.org/abs/gr-qc/0305096}{{\ttfamily arXiv:gr-qc/0305096}}.

\bibitem{Tashiro:2003qp}
H.~Tashiro, T.~Chiba, and M.~Sasaki, ``{Reheating after quintessential
  inflation and gravitational waves},''
  \href{http://dx.doi.org/10.1088/0264-9381/21/7/004}{{\em Class. Quant. Grav.}
  {\bfseries 21} (2004) 1761--1772},
  \href{http://arxiv.org/abs/gr-qc/0307068}{{\ttfamily arXiv:gr-qc/0307068}}.

\bibitem{Nakayama:2008ip}
K.~Nakayama, S.~Saito, Y.~Suwa, and J.~Yokoyama, ``{Space laser interferometers
  can determine the thermal history of the early Universe},''
  \href{http://dx.doi.org/10.1103/PhysRevD.77.124001}{{\em Phys. Rev. D}
  {\bfseries 77} (2008) 124001},
  \href{http://arxiv.org/abs/0802.2452}{{\ttfamily arXiv:0802.2452 [hep-ph]}}.

\bibitem{Nakayama:2008wy}
K.~Nakayama, S.~Saito, Y.~Suwa, and J.~Yokoyama, ``{Probing reheating
  temperature of the universe with gravitational wave background},''
  \href{http://dx.doi.org/10.1088/1475-7516/2008/06/020}{{\em JCAP} {\bfseries
  06} (2008) 020}, \href{http://arxiv.org/abs/0804.1827}{{\ttfamily
  arXiv:0804.1827 [astro-ph]}}.

\bibitem{Kuroyanagi:2008ye}
S.~Kuroyanagi, T.~Chiba, and N.~Sugiyama, ``{Precision calculations of the
  gravitational wave background spectrum from inflation},''
  \href{http://dx.doi.org/10.1103/PhysRevD.79.103501}{{\em Phys. Rev. D}
  {\bfseries 79} (2009) 103501},
  \href{http://arxiv.org/abs/0804.3249}{{\ttfamily arXiv:0804.3249
  [astro-ph]}}.

\bibitem{Mukohyama:2009zs}
S.~Mukohyama, K.~Nakayama, F.~Takahashi, and S.~Yokoyama, ``{Phenomenological
  Aspects of Horava-Lifshitz Cosmology},''
  \href{http://dx.doi.org/10.1016/j.physletb.2009.07.005}{{\em Phys. Lett. B}
  {\bfseries 679} (2009) 6--9},
  \href{http://arxiv.org/abs/0905.0055}{{\ttfamily arXiv:0905.0055 [hep-th]}}.

\bibitem{Nakayama:2009ce}
K.~Nakayama and J.~Yokoyama, ``{Gravitational Wave Background and
  Non-Gaussianity as a Probe of the Curvaton Scenario},''
  \href{http://dx.doi.org/10.1088/1475-7516/2010/01/010}{{\em JCAP} {\bfseries
  01} (2010) 010}, \href{http://arxiv.org/abs/0910.0715}{{\ttfamily
  arXiv:0910.0715 [astro-ph.CO]}}.

\bibitem{Durrer:2011bi}
R.~Durrer and J.~Hasenkamp, ``{Testing Superstring Theories with Gravitational
  Waves},'' \href{http://dx.doi.org/10.1103/PhysRevD.84.064027}{{\em Phys. Rev.
  D} {\bfseries 84} (2011) 064027},
  \href{http://arxiv.org/abs/1105.5283}{{\ttfamily arXiv:1105.5283 [gr-qc]}}.

\bibitem{Jinno:2011sw}
R.~Jinno, T.~Moroi, and K.~Nakayama, ``{Imprints of Cosmic Phase Transition in
  Inflationary Gravitational Waves},''
  \href{http://dx.doi.org/10.1016/j.physletb.2012.05.061}{{\em Phys. Lett. B}
  {\bfseries 713} (2012) 129--132},
  \href{http://arxiv.org/abs/1112.0084}{{\ttfamily arXiv:1112.0084 [hep-ph]}}.

\bibitem{Kuroyanagi:2011fy}
S.~Kuroyanagi, K.~Nakayama, and S.~Saito, ``{Prospects for determination of
  thermal history after inflation with future gravitational wave detectors},''
  \href{http://dx.doi.org/10.1103/PhysRevD.84.123513}{{\em Phys. Rev. D}
  {\bfseries 84} (2011) 123513},
  \href{http://arxiv.org/abs/1110.4169}{{\ttfamily arXiv:1110.4169
  [astro-ph.CO]}}.

\bibitem{Jinno:2012xb}
R.~Jinno, T.~Moroi, and K.~Nakayama, ``{Probing dark radiation with
  inflationary gravitational waves},''
  \href{http://dx.doi.org/10.1103/PhysRevD.86.123502}{{\em Phys. Rev. D}
  {\bfseries 86} (2012) 123502},
  \href{http://arxiv.org/abs/1208.0184}{{\ttfamily arXiv:1208.0184
  [astro-ph.CO]}}.

\bibitem{Jinno:2013xqa}
R.~Jinno, T.~Moroi, and K.~Nakayama, ``{Inflationary Gravitational Waves and
  the Evolution of the Early Universe},''
  \href{http://dx.doi.org/10.1088/1475-7516/2014/01/040}{{\em JCAP} {\bfseries
  01} (2014) 040}, \href{http://arxiv.org/abs/1307.3010}{{\ttfamily
  arXiv:1307.3010 [hep-ph]}}.

\bibitem{Jinno:2014qka}
R.~Jinno, T.~Moroi, and T.~Takahashi, ``{Studying Inflation with Future
  Space-Based Gravitational Wave Detectors},''
  \href{http://dx.doi.org/10.1088/1475-7516/2014/12/006}{{\em JCAP} {\bfseries
  12} (2014) 006}, \href{http://arxiv.org/abs/1406.1666}{{\ttfamily
  arXiv:1406.1666 [astro-ph.CO]}}.

\bibitem{Kuroyanagi:2014qza}
S.~Kuroyanagi, K.~Nakayama, and J.~Yokoyama, ``{Prospects of determination of
  reheating temperature after inflation by DECIGO},''
  \href{http://dx.doi.org/10.1093/ptep/ptu176}{{\em PTEP} {\bfseries 2015}
  no.~1, (2015) 013E02}, \href{http://arxiv.org/abs/1410.6618}{{\ttfamily
  arXiv:1410.6618 [astro-ph.CO]}}.

\bibitem{Minami:2025waa}
K.~Minami, K.~Mukaida, and K.~Nakayama, ``{Reheating with Thermal Dissipation
  and Primordial Gravitational Waves},''
  \href{http://arxiv.org/abs/2510.02481}{{\ttfamily arXiv:2510.02481
  [astro-ph.CO]}}.

\bibitem{Ema:2015dka}
Y.~Ema, R.~Jinno, K.~Mukaida, and K.~Nakayama, ``{Gravitational Effects on
  Inflaton Decay},''
  \href{http://dx.doi.org/10.1088/1475-7516/2015/05/038}{{\em JCAP} {\bfseries
  05} (2015) 038}, \href{http://arxiv.org/abs/1502.02475}{{\ttfamily
  arXiv:1502.02475 [hep-ph]}}.

\bibitem{Ema:2016hlw}
Y.~Ema, R.~Jinno, K.~Mukaida, and K.~Nakayama, ``{Gravitational particle
  production in oscillating backgrounds and its cosmological implications},''
  \href{http://dx.doi.org/10.1103/PhysRevD.94.063517}{{\em Phys. Rev. D}
  {\bfseries 94} no.~6, (2016) 063517},
  \href{http://arxiv.org/abs/1604.08898}{{\ttfamily arXiv:1604.08898
  [hep-ph]}}.

\bibitem{Ema:2020ggo}
Y.~Ema, R.~Jinno, and K.~Nakayama, ``{High-frequency Graviton from Inflaton
  Oscillation},'' \href{http://dx.doi.org/10.1088/1475-7516/2020/09/015}{{\em
  JCAP} {\bfseries 09} (2020) 015},
  \href{http://arxiv.org/abs/2006.09972}{{\ttfamily arXiv:2006.09972
  [astro-ph.CO]}}.

\bibitem{Khlebnikov:1997di}
S.~Y. Khlebnikov and I.~I. Tkachev, ``{Relic gravitational waves produced after
  preheating},'' \href{http://dx.doi.org/10.1103/PhysRevD.56.653}{{\em Phys.
  Rev. D} {\bfseries 56} (1997) 653--660},
  \href{http://arxiv.org/abs/hep-ph/9701423}{{\ttfamily arXiv:hep-ph/9701423}}.

\bibitem{Easther:2006vd}
R.~Easther, J.~T. Giblin, Jr., and E.~A. Lim, ``{Gravitational Wave Production
  At The End Of Inflation},''
  \href{http://dx.doi.org/10.1103/PhysRevLett.99.221301}{{\em Phys. Rev. Lett.}
  {\bfseries 99} (2007) 221301},
  \href{http://arxiv.org/abs/astro-ph/0612294}{{\ttfamily
  arXiv:astro-ph/0612294}}.

\bibitem{Easther:2006gt}
R.~Easther and E.~A. Lim, ``{Stochastic gravitational wave production after
  inflation},'' \href{http://dx.doi.org/10.1088/1475-7516/2006/04/010}{{\em
  JCAP} {\bfseries 04} (2006) 010},
  \href{http://arxiv.org/abs/astro-ph/0601617}{{\ttfamily
  arXiv:astro-ph/0601617}}.

\bibitem{Garcia-Bellido:2007nns}
J.~Garcia-Bellido and D.~G. Figueroa, ``{A stochastic background of
  gravitational waves from hybrid preheating},''
  \href{http://dx.doi.org/10.1103/PhysRevLett.98.061302}{{\em Phys. Rev. Lett.}
  {\bfseries 98} (2007) 061302},
  \href{http://arxiv.org/abs/astro-ph/0701014}{{\ttfamily
  arXiv:astro-ph/0701014}}.

\bibitem{Garcia-Bellido:2007fiu}
J.~Garcia-Bellido, D.~G. Figueroa, and A.~Sastre, ``{A Gravitational Wave
  Background from Reheating after Hybrid Inflation},''
  \href{http://dx.doi.org/10.1103/PhysRevD.77.043517}{{\em Phys. Rev. D}
  {\bfseries 77} (2008) 043517},
  \href{http://arxiv.org/abs/0707.0839}{{\ttfamily arXiv:0707.0839 [hep-ph]}}.

\bibitem{Dufaux:2007pt}
J.~F. Dufaux, A.~Bergman, G.~N. Felder, L.~Kofman, and J.-P. Uzan, ``{Theory
  and Numerics of Gravitational Waves from Preheating after Inflation},''
  \href{http://dx.doi.org/10.1103/PhysRevD.76.123517}{{\em Phys. Rev. D}
  {\bfseries 76} (2007) 123517},
  \href{http://arxiv.org/abs/0707.0875}{{\ttfamily arXiv:0707.0875
  [astro-ph]}}.

\bibitem{Dufaux:2008dn}
J.-F. Dufaux, G.~Felder, L.~Kofman, and O.~Navros, ``{Gravity Waves from
  Tachyonic Preheating after Hybrid Inflation},''
  \href{http://dx.doi.org/10.1088/1475-7516/2009/03/001}{{\em JCAP} {\bfseries
  03} (2009) 001}, \href{http://arxiv.org/abs/0812.2917}{{\ttfamily
  arXiv:0812.2917 [astro-ph]}}.

\bibitem{Dolgov:1989us}
A.~D. Dolgov and D.~P. Kirilova, ``{ON PARTICLE CREATION BY A TIME DEPENDENT
  SCALAR FIELD},'' {\em Sov. J. Nucl. Phys.} {\bfseries 51} (1990) 172--177.

\bibitem{Traschen:1990sw}
J.~H. Traschen and R.~H. Brandenberger, ``{Particle Production During
  Out-of-equilibrium Phase Transitions},''
  \href{http://dx.doi.org/10.1103/PhysRevD.42.2491}{{\em Phys. Rev. D}
  {\bfseries 42} (1990) 2491--2504}.

\bibitem{Shtanov:1994ce}
Y.~Shtanov, J.~H. Traschen, and R.~H. Brandenberger, ``{Universe reheating
  after inflation},'' \href{http://dx.doi.org/10.1103/PhysRevD.51.5438}{{\em
  Phys. Rev. D} {\bfseries 51} (1995) 5438--5455},
  \href{http://arxiv.org/abs/hep-ph/9407247}{{\ttfamily arXiv:hep-ph/9407247}}.

\bibitem{Kofman:1994rk}
L.~Kofman, A.~D. Linde, and A.~A. Starobinsky, ``{Reheating after inflation},''
  \href{http://dx.doi.org/10.1103/PhysRevLett.73.3195}{{\em Phys. Rev. Lett.}
  {\bfseries 73} (1994) 3195--3198},
  \href{http://arxiv.org/abs/hep-th/9405187}{{\ttfamily arXiv:hep-th/9405187}}.

\bibitem{Kofman:1997yn}
L.~Kofman, A.~D. Linde, and A.~A. Starobinsky, ``{Towards the theory of
  reheating after inflation},''
  \href{http://dx.doi.org/10.1103/PhysRevD.56.3258}{{\em Phys. Rev. D}
  {\bfseries 56} (1997) 3258--3295},
  \href{http://arxiv.org/abs/hep-ph/9704452}{{\ttfamily arXiv:hep-ph/9704452}}.

\bibitem{Choi:2024ilx}
G.~Choi, W.~Ke, and K.~A. Olive, ``{Minimal production of prompt gravitational
  waves during reheating},''
  \href{http://dx.doi.org/10.1103/PhysRevD.109.083516}{{\em Phys. Rev. D}
  {\bfseries 109} no.~8, (2024) 083516},
  \href{http://arxiv.org/abs/2402.04310}{{\ttfamily arXiv:2402.04310
  [hep-ph]}}.

\bibitem{Xu:2024fjl}
Y.~Xu, ``{Ultra-high frequency gravitational waves from scattering,
  Bremsstrahlung and decay during reheating},''
  \href{http://dx.doi.org/10.1007/JHEP10(2024)174}{{\em JHEP} {\bfseries 10}
  (2024) 174}, \href{http://arxiv.org/abs/2407.03256}{{\ttfamily
  arXiv:2407.03256 [hep-ph]}}.

\bibitem{Bernal:2025lxp}
N.~Bernal, Q.-f. Wu, X.-J. Xu, and Y.~Xu, ``{Pre-thermalized gravitational
  waves},'' \href{http://dx.doi.org/10.1007/JHEP08(2025)125}{{\em JHEP}
  {\bfseries 08} (2025) 125}, \href{http://arxiv.org/abs/2503.10756}{{\ttfamily
  arXiv:2503.10756 [hep-ph]}}.

\bibitem{Xu:2025wjq}
X.-J. Xu, Y.~Xu, Q.~Yin, and J.~Zhu, ``{Full-spectrum analysis of gravitational
  wave production from inflation to reheating},''
  \href{http://dx.doi.org/10.1007/JHEP10(2025)141}{{\em JHEP} {\bfseries 10}
  (2025) 141}, \href{http://arxiv.org/abs/2505.08868}{{\ttfamily
  arXiv:2505.08868 [hep-ph]}}.

\bibitem{Birrell:1982ix}
N.~D. Birrell and P.~C.~W. Davies,
  \href{http://dx.doi.org/10.1017/CBO9780511622632}{{\em {Quantum Fields in
  Curved Space}}}.
\newblock Cambridge Monographs on Mathematical Physics. Cambridge University
  Press, Cambridge, UK, 1982.

\bibitem{Ema:2018ucl}
Y.~Ema, K.~Nakayama, and Y.~Tang, ``{Production of Purely Gravitational Dark
  Matter},'' \href{http://dx.doi.org/10.1007/JHEP09(2018)135}{{\em JHEP}
  {\bfseries 09} (2018) 135}, \href{http://arxiv.org/abs/1804.07471}{{\ttfamily
  arXiv:1804.07471 [hep-ph]}}.

\bibitem{Liddle:2000cg}
A.~R. Liddle and D.~H. Lyth,
  \href{http://dx.doi.org/10.1017/CBO9781139175180}{{\em {Cosmological
  inflation and large scale structure}}}.
\newblock 2000.

\bibitem{Lozanov:2016hid}
K.~D. Lozanov and M.~A. Amin, ``{Equation of State and Duration to Radiation
  Domination after Inflation},''
  \href{http://dx.doi.org/10.1103/PhysRevLett.119.061301}{{\em Phys. Rev.
  Lett.} {\bfseries 119} no.~6, (2017) 061301},
  \href{http://arxiv.org/abs/1608.01213}{{\ttfamily arXiv:1608.01213
  [astro-ph.CO]}}.

\bibitem{Lozanov:2017hjm}
K.~D. Lozanov and M.~A. Amin, ``{Self-resonance after inflation: oscillons,
  transients and radiation domination},''
  \href{http://dx.doi.org/10.1103/PhysRevD.97.023533}{{\em Phys. Rev. D}
  {\bfseries 97} no.~2, (2018) 023533},
  \href{http://arxiv.org/abs/1710.06851}{{\ttfamily arXiv:1710.06851
  [astro-ph.CO]}}.

\bibitem{Antusch:2021aiw}
S.~Antusch, D.~G. Figueroa, K.~Marschall, and F.~Torrenti, ``{Characterizing
  the postinflationary reheating history: Single daughter field with
  quadratic-quadratic interaction},''
  \href{http://dx.doi.org/10.1103/PhysRevD.105.043532}{{\em Phys. Rev. D}
  {\bfseries 105} no.~4, (2022) 043532},
  \href{http://arxiv.org/abs/2112.11280}{{\ttfamily arXiv:2112.11280
  [astro-ph.CO]}}.

\bibitem{Garcia:2023dyf}
M.~A.~G. Garcia, M.~Gross, Y.~Mambrini, K.~A. Olive, M.~Pierre, and J.-H. Yoon,
  ``{Effects of fragmentation on post-inflationary reheating},''
  \href{http://dx.doi.org/10.1088/1475-7516/2023/12/028}{{\em JCAP} {\bfseries
  12} (2023) 028}, \href{http://arxiv.org/abs/2308.16231}{{\ttfamily
  arXiv:2308.16231 [hep-ph]}}.

\bibitem{Eroncel:2025bcb}
C.~Er{\"o}ncel, Y.~Gouttenoire, R.~Sato, G.~Servant, and P.~Simakachorn,
  ``{Universal Bound on the Duration of a Kination Era},''
  \href{http://dx.doi.org/10.1103/k7ty-gwjg}{{\em Phys. Rev. Lett.} {\bfseries
  135} no.~10, (2025) 101002},
  \href{http://arxiv.org/abs/2501.17226}{{\ttfamily arXiv:2501.17226
  [hep-ph]}}.

\bibitem{Planck:2018vyg}
{\bfseries Planck} Collaboration, N.~Aghanim {\em et~al.}, ``{Planck 2018
  results. VI. Cosmological parameters},''
  \href{http://dx.doi.org/10.1051/0004-6361/201833910}{{\em Astron. Astrophys.}
  {\bfseries 641} (2020) A6}, \href{http://arxiv.org/abs/1807.06209}{{\ttfamily
  arXiv:1807.06209 [astro-ph.CO]}}. [Erratum: Astron.Astrophys. 652, C4
  (2021)].

\bibitem{Nakayama:2013jka}
K.~Nakayama, F.~Takahashi, and T.~T. Yanagida, ``{Polynomial Chaotic Inflation
  in the Planck Era},''
  \href{http://dx.doi.org/10.1016/j.physletb.2013.06.050}{{\em Phys. Lett. B}
  {\bfseries 725} (2013) 111--114},
  \href{http://arxiv.org/abs/1303.7315}{{\ttfamily arXiv:1303.7315 [hep-ph]}}.

\bibitem{Nakayama:2013txa}
K.~Nakayama, F.~Takahashi, and T.~T. Yanagida, ``{Polynomial Chaotic Inflation
  in Supergravity},''
  \href{http://dx.doi.org/10.1088/1475-7516/2013/08/038}{{\em JCAP} {\bfseries
  08} (2013) 038}, \href{http://arxiv.org/abs/1305.5099}{{\ttfamily
  arXiv:1305.5099 [hep-ph]}}.

\bibitem{Linde:1983gd}
A.~D. Linde, ``{Chaotic Inflation},''
  \href{http://dx.doi.org/10.1016/0370-2693(83)90837-7}{{\em Phys. Lett. B}
  {\bfseries 129} (1983) 177--181}.

\bibitem{Lyth:2009imm}
D.~H. Lyth and A.~R. Liddle,
  \href{http://dx.doi.org/10.1017/cbo9780511819209}{{\em {The Primordial
  Density Perturbation}}}.
\newblock 6, 2009.

\bibitem{Gorbunov:2010bn}
D.~S. Gorbunov and A.~G. Panin, ``{Scalaron the mighty: producing dark matter
  and baryon asymmetry at reheating},''
  \href{http://dx.doi.org/10.1016/j.physletb.2011.04.067}{{\em Phys. Lett. B}
  {\bfseries 700} (2011) 157--162},
  \href{http://arxiv.org/abs/1009.2448}{{\ttfamily arXiv:1009.2448 [hep-ph]}}.

\bibitem{Li:2021fao}
Q.~Li, T.~Moroi, K.~Nakayama, and W.~Yin, ``{Hidden dark matter from
  Starobinsky inflation},''
  \href{http://dx.doi.org/10.1007/JHEP09(2021)179}{{\em JHEP} {\bfseries 09}
  (2021) 179}, \href{http://arxiv.org/abs/2105.13358}{{\ttfamily
  arXiv:2105.13358 [hep-ph]}}.

\bibitem{Kumekawa:1994gx}
K.~Kumekawa, T.~Moroi, and T.~Yanagida, ``{Flat potential for inflaton with a
  discrete R invariance in supergravity},''
  \href{http://dx.doi.org/10.1143/PTP.92.437}{{\em Prog. Theor. Phys.}
  {\bfseries 92} (1994) 437--448},
  \href{http://arxiv.org/abs/hep-ph/9405337}{{\ttfamily arXiv:hep-ph/9405337}}.

\bibitem{Izawa:1996dv}
K.~I. Izawa and T.~Yanagida, ``{Natural new inflation in broken
  supergravity},'' \href{http://dx.doi.org/10.1016/S0370-2693(96)01638-3}{{\em
  Phys. Lett. B} {\bfseries 393} (1997) 331--336},
  \href{http://arxiv.org/abs/hep-ph/9608359}{{\ttfamily arXiv:hep-ph/9608359}}.

\bibitem{Asaka:1999jb}
T.~Asaka, K.~Hamaguchi, M.~Kawasaki, and T.~Yanagida, ``{Leptogenesis in
  inflationary universe},''
  \href{http://dx.doi.org/10.1103/PhysRevD.61.083512}{{\em Phys. Rev. D}
  {\bfseries 61} (2000) 083512},
  \href{http://arxiv.org/abs/hep-ph/9907559}{{\ttfamily arXiv:hep-ph/9907559}}.

\bibitem{Senoguz:2004ky}
V.~N. Senoguz and Q.~Shafi, ``{New inflation, preinflation, and
  leptogenesis},'' \href{http://dx.doi.org/10.1016/j.physletb.2004.05.077}{{\em
  Phys. Lett. B} {\bfseries 596} (2004) 8--15},
  \href{http://arxiv.org/abs/hep-ph/0403294}{{\ttfamily arXiv:hep-ph/0403294}}.

\bibitem{Kohri:2007gq}
K.~Kohri, C.-M. Lin, and D.~H. Lyth, ``{More hilltop inflation models},''
  \href{http://dx.doi.org/10.1088/1475-7516/2007/12/004}{{\em JCAP} {\bfseries
  12} (2007) 004}, \href{http://arxiv.org/abs/0707.3826}{{\ttfamily
  arXiv:0707.3826 [hep-ph]}}.

\bibitem{Nakayama:2012dw}
K.~Nakayama and F.~Takahashi, ``{PeV-scale Supersymmetry from New Inflation},''
  \href{http://dx.doi.org/10.1088/1475-7516/2012/05/035}{{\em JCAP} {\bfseries
  05} (2012) 035}, \href{http://arxiv.org/abs/1203.0323}{{\ttfamily
  arXiv:1203.0323 [hep-ph]}}.

\bibitem{Ema:2017rkk}
Y.~Ema, K.~Mukaida, and K.~Nakayama, ``{Electroweak Vacuum Metastability and
  Low-scale Inflation},''
  \href{http://dx.doi.org/10.1088/1475-7516/2017/12/030}{{\em JCAP} {\bfseries
  12} (2017) 030}, \href{http://arxiv.org/abs/1706.08920}{{\ttfamily
  arXiv:1706.08920 [hep-ph]}}.

\bibitem{Brax:2010ai}
P.~Brax, J.-F. Dufaux, and S.~Mariadassou, ``{Preheating after Small-Field
  Inflation},'' \href{http://dx.doi.org/10.1103/PhysRevD.83.103510}{{\em Phys.
  Rev. D} {\bfseries 83} (2011) 103510},
  \href{http://arxiv.org/abs/1012.4656}{{\ttfamily arXiv:1012.4656 [hep-th]}}.

\bibitem{Antusch:2015nla}
S.~Antusch, D.~Nolde, and S.~Orani, ``{Hill crossing during preheating after
  hilltop inflation},''
  \href{http://dx.doi.org/10.1088/1475-7516/2015/06/009}{{\em JCAP} {\bfseries
  06} (2015) 009}, \href{http://arxiv.org/abs/1503.06075}{{\ttfamily
  arXiv:1503.06075 [hep-ph]}}.

\bibitem{Antusch:2015vna}
S.~Antusch, F.~Cefala, D.~Nolde, and S.~Orani, ``{Parametric resonance after
  hilltop inflation caused by an inhomogeneous inflaton field},''
  \href{http://dx.doi.org/10.1088/1475-7516/2016/02/044}{{\em JCAP} {\bfseries
  02} (2016) 044}, \href{http://arxiv.org/abs/1510.04856}{{\ttfamily
  arXiv:1510.04856 [hep-ph]}}.

\bibitem{Kallosh:2013hoa}
R.~Kallosh and A.~Linde, ``{Universality Class in Conformal Inflation},''
  \href{http://dx.doi.org/10.1088/1475-7516/2013/07/002}{{\em JCAP} {\bfseries
  07} (2013) 002}, \href{http://arxiv.org/abs/1306.5220}{{\ttfamily
  arXiv:1306.5220 [hep-th]}}.

\bibitem{Kallosh:2013yoa}
R.~Kallosh, A.~Linde, and D.~Roest, ``{Superconformal Inflationary
  $\alpha$-Attractors},'' \href{http://dx.doi.org/10.1007/JHEP11(2013)198}{{\em
  JHEP} {\bfseries 11} (2013) 198},
  \href{http://arxiv.org/abs/1311.0472}{{\ttfamily arXiv:1311.0472 [hep-th]}}.

\bibitem{Galante:2014ifa}
M.~Galante, R.~Kallosh, A.~Linde, and D.~Roest, ``{Unity of Cosmological
  Inflation Attractors},''
  \href{http://dx.doi.org/10.1103/PhysRevLett.114.141302}{{\em Phys. Rev.
  Lett.} {\bfseries 114} no.~14, (2015) 141302},
  \href{http://arxiv.org/abs/1412.3797}{{\ttfamily arXiv:1412.3797 [hep-th]}}.

\bibitem{Nakayama:2018ptw}
K.~Nakayama and Y.~Tang, ``{Stochastic Gravitational Waves from Particle
  Origin},'' \href{http://dx.doi.org/10.1016/j.physletb.2018.11.023}{{\em Phys.
  Lett. B} {\bfseries 788} (2019) 341--346},
  \href{http://arxiv.org/abs/1810.04975}{{\ttfamily arXiv:1810.04975
  [hep-ph]}}. [Erratum: Phys.Lett.B 839, 137787 (2023)].

\bibitem{Barman:2023ymn}
B.~Barman, N.~Bernal, Y.~Xu, and {\'O}.~Zapata, ``{Gravitational wave from
  graviton Bremsstrahlung during reheating},''
  \href{http://dx.doi.org/10.1088/1475-7516/2023/05/019}{{\em JCAP} {\bfseries
  05} (2023) 019}, \href{http://arxiv.org/abs/2301.11345}{{\ttfamily
  arXiv:2301.11345 [hep-ph]}}.

\bibitem{Barman:2023rpg}
B.~Barman, N.~Bernal, Y.~Xu, and {\'O}.~Zapata, ``{Bremsstrahlung-induced
  gravitational waves in monomial potentials during reheating},''
  \href{http://dx.doi.org/10.1103/PhysRevD.108.083524}{{\em Phys. Rev. D}
  {\bfseries 108} no.~8, (2023) 083524},
  \href{http://arxiv.org/abs/2305.16388}{{\ttfamily arXiv:2305.16388
  [hep-ph]}}.

\bibitem{Hu:2024awd}
W.~Hu, K.~Nakayama, V.~Takhistov, and Y.~Tang, ``{Gravitational wave probe of
  Planck-scale physics after inflation},''
  \href{http://dx.doi.org/10.1016/j.physletb.2024.138958}{{\em Phys. Lett. B}
  {\bfseries 856} (2024) 138958},
  \href{http://arxiv.org/abs/2403.13882}{{\ttfamily arXiv:2403.13882
  [hep-ph]}}.

\bibitem{Hu:2024bha}
W.-Y. Hu, K.~Nakayama, V.~Takhistov, and Y.~Tang, ``{Dual gravitational wave
  signatures of instant preheating},''
  \href{http://dx.doi.org/10.1088/1475-7516/2025/01/029}{{\em JCAP} {\bfseries
  01} (2025) 029}, \href{http://arxiv.org/abs/2409.06483}{{\ttfamily
  arXiv:2409.06483 [astro-ph.CO]}}.

\bibitem{Ema:2021fdz}
Y.~Ema, K.~Mukaida, and K.~Nakayama, ``{Scalar field couplings to quadratic
  curvature and decay into gravitons},''
  \href{http://dx.doi.org/10.1007/JHEP05(2022)087}{{\em JHEP} {\bfseries 05}
  (2022) 087}, \href{http://arxiv.org/abs/2112.12774}{{\ttfamily
  arXiv:2112.12774 [hep-ph]}}.

\bibitem{Mudrunka:2023wxy}
K.~Mudrunka and K.~Nakayama, ``{Probing Gauss-Bonnet-corrected inflation with
  gravitational waves},''
  \href{http://dx.doi.org/10.1088/1475-7516/2024/05/069}{{\em JCAP} {\bfseries
  05} (2024) 069}, \href{http://arxiv.org/abs/2312.15766}{{\ttfamily
  arXiv:2312.15766 [astro-ph.CO]}}.

\bibitem{Tokareva:2023mrt}
A.~Tokareva, ``{Gravitational waves from inflaton decay and bremsstrahlung},''
  \href{http://dx.doi.org/10.1016/j.physletb.2024.138695}{{\em Phys. Lett. B}
  {\bfseries 853} (2024) 138695},
  \href{http://arxiv.org/abs/2312.16691}{{\ttfamily arXiv:2312.16691
  [hep-ph]}}.

\bibitem{Strumia:2025dfn}
A.~Strumia and G.~Landini, ``{Optical gravitational waves as signals of
  gravitationally-decaying particles},''
  \href{http://dx.doi.org/10.1007/JHEP04(2025)068}{{\em JHEP} {\bfseries 04}
  (2025) 068}, \href{http://arxiv.org/abs/2501.09794}{{\ttfamily
  arXiv:2501.09794 [hep-ph]}}.

\bibitem{Nakayama:2025xkn}
K.~Nakayama, F.~Takahashi, and J.~Wada, ``{Gravitational Decays of Secluded
  Scalars and Graviton Dark Radiation},''
  \href{http://arxiv.org/abs/2512.03662}{{\ttfamily arXiv:2512.03662
  [hep-ph]}}.

\bibitem{Ghiglieri:2015nfa}
J.~Ghiglieri and M.~Laine, ``{Gravitational wave background from Standard Model
  physics: Qualitative features},''
  \href{http://dx.doi.org/10.1088/1475-7516/2015/07/022}{{\em JCAP} {\bfseries
  07} (2015) 022}, \href{http://arxiv.org/abs/1504.02569}{{\ttfamily
  arXiv:1504.02569 [hep-ph]}}.

\bibitem{Ghiglieri:2020mhm}
J.~Ghiglieri, G.~Jackson, M.~Laine, and Y.~Zhu, ``{Gravitational wave
  background from Standard Model physics: Complete leading order},''
  \href{http://dx.doi.org/10.1007/JHEP07(2020)092}{{\em JHEP} {\bfseries 07}
  (2020) 092}, \href{http://arxiv.org/abs/2004.11392}{{\ttfamily
  arXiv:2004.11392 [hep-ph]}}.

\bibitem{Ringwald:2020ist}
A.~Ringwald, J.~Sch{\"u}tte-Engel, and C.~Tamarit, ``{Gravitational Waves as a
  Big Bang Thermometer},''
  \href{http://dx.doi.org/10.1088/1475-7516/2021/03/054}{{\em JCAP} {\bfseries
  03} (2021) 054}, \href{http://arxiv.org/abs/2011.04731}{{\ttfamily
  arXiv:2011.04731 [hep-ph]}}.

\bibitem{Ghiglieri:2022rfp}
J.~Ghiglieri, J.~Sch{\"u}tte-Engel, and E.~Speranza, ``{Freezing-in
  gravitational waves},''
  \href{http://dx.doi.org/10.1103/PhysRevD.109.023538}{{\em Phys. Rev. D}
  {\bfseries 109} no.~2, (2024) 023538},
  \href{http://arxiv.org/abs/2211.16513}{{\ttfamily arXiv:2211.16513
  [hep-ph]}}.

\bibitem{Baumann:2022mni}
D.~Baumann, \href{http://dx.doi.org/10.1017/9781108937092}{{\em {Cosmology}}}.
\newblock Cambridge University Press, 7, 2022.

\end{thebibliography}\endgroup

\end{document}